\documentclass{article}

\usepackage{graphicx}

\usepackage{amsmath,amssymb}

\begin{document}
\begin{titlepage}

\begin{flushright}
BUTP/2001-16\\
OUTP-01-45P\\
\end{flushright}
\vspace{1cm}
\begin{center}
{\LARGE Classically Perfect Gauge Actions on Anisotropic Lattices}
\footnote{Work supported in part by Schweizerischer Nationalfonds}

\vspace{1cm}
{\large
Philipp R\"ufenacht}
\\
\vspace{0.5cm}
Institute for Theoretical Physics, University of Bern\\
Sidlerstrasse 5, CH-3012 Bern, Switzerland
\\
\vspace{1cm}
{\large 
Urs Wenger}
\\
\vspace{0.5cm}
Theoretical Physics, Oxford University\\
 1 Keble Road, Oxford OX1 3NP, United Kingdom
\\\vspace{1cm}
{\bf Abstract}
\end{center}
\vspace{5mm}

\begin{quote}
We present a method for constructing classically perfect anisotropic
actions for SU(3) gauge theory based on an isotropic Fixed Point
Action. The action is pa\-ra\-metri\-sed using smeared (``fat'')
links. The construction is done explicitly for anisotropy
$\xi=a_s/a_t=2$ and $4$. The corresponding renormalised
anisotropies are determined using the torelon dispersion relation. The
renormalisation of the anisotropy is small and the parametrisation
describes the true action well. Quantities such as the static
quark-antiquark potential, the critical temperature of the deconfining
phase transition and the low-lying glueball spectrum are measured on
lattices with anisotropy $\xi=2$. The mass of the scalar $0^{++}$
glueball is determined to be 1580(60)~MeV, while the tensor $2^{++}$
glueball is at 2430(60)~MeV.
\end{quote}

\vfill
\end{titlepage}
\section{Introduction}\label{intro}

In Lattice QCD the energy of a physical state is measured by studying the
decay of correlators of operators which have a non-zero transition matrix
element between the vacuum and the state under consideration. If the state is heavy,
these correlators decay very fast in Euclidean time and one has to make sure
that the temporal lattice spacing is small enough such that the signal
of the correlator can be accurately traced over a few time slices
before it disappears in the statistical noise. At the same time, one
has to pay attention to choose the physical volume of the lattice
large enough so that there are no significant finite-size
effects. Both these requirements together lead to lattices with a
large number of lattice sites and hence to computationally
very expensive simulations.  The obvious way out of
this dilemma is to use a smaller lattice spacing in temporal direction
compared to the spatial directions, i.e.~the use of anisotropic lattices.

Anisotropic lattices have been widely used over the last few
years. Studies comprising excited states of nucleons
\cite{Lee:2000hh}, heavy-quark bound states
\cite{AliKhan:2000bv,Chen:2000ej,Chen:2000qj}, heavy-light mesons
\cite{Lewis:1999yw,Lewis:2000sv}, heavy meson semi-leptonic decays
\cite{Shigemitsu:2000hj}, long range properties and excited states of
the quark-antiquark potential
\cite{Darmohval:1999ht,Bali:1999hx,Deldar:1999id,Deldar:1999vi} as
well as states composed purely of gluons (glueballs)
\cite{Morningstar:1997ff,Morningstar:1999rf,Morningstar:1999dh,Liu:2000ce}
or gluons and quarks (hybrids)
\cite{Manke:1998qc,AliKhan:1999yz,Manke:1999ru,Drummond:1998fd,Drummond:1999db,Juge:1999aw},
have been performed, mainly using the standard anisotropic Wilson
discretisation or a mean-link and Symanzik improved anisotropic action
\cite{Morningstar:1997ze}.  In fact, improved actions turned out to be
crucial in anisotropic lattice calculations due to the coarse spatial
lattice spacings employed. The discretisation of the
continuum action, however, can be done in many different ways leading to
different improvement schemes. A radical approach of improving lattice
actions, based on Wilson's Renormalisation Group ideas, has been
suggested by Hasenfratz and Niedermayer \cite{Hasenfratz:1994sp},
namely the use of actions that are classically perfect, i.e.~there are
no lattice artifacts on the solutions of the lattice equations of
motion. For SU(3) gauge theory the classically perfect FP action has
been constructed and tested in
\cite{DeGrand:1995ji,DeGrand:1995jk,DeGrand:1996ab,Blatter:1996ti,Niedermayer:2000yx}
and the ansatz has been extended to include FP actions for fermions as
well
\cite{Bietenholz:1996cy,Bietenholz:1996qc,DeGrand:1997nc,Hasenfratz:2000xz,Hasenfratz:2000qb}. In
the case of SU(2) gauge theory the FP action has been constructed in
\cite{DeGrand:1996ih,DeGrand:1996zb,DeGrand:1997gu} and its classical
properties have been tested on classical instanton solutions, both in
SU(2) and SU(3) \cite{Farchioni_Papa:98}.  A classically perfect gauge
action on anisotropic lattices, however, has so far been absent.

It is thus the goal of this work to fill the gap and to present the
construction of classically perfect gauge actions on anisotropic lattices. The
method has been introduced in \cite{Rufenacht:2000px} and it is presented in
more detail in \cite{Rufenacht:2001aa}. It relies upon a recent
parametrisation of the isotropic classically perfect Fixed Point (FP) action 
\cite{Niedermayer:2000yx,Wenger:2000aa,Niedermayer:2000ts} that includes a
rich structure of operators as it is based on plaquettes built from simple
gauge links as well as from smeared ("fat") links. We explicitly
construct the $\xi=2$ and 4 actions on coarse configurations typically
occurring in Monte Carlo (MC) simulations.  The properties of the $\xi=2$
action are studied by performing measurements of the torelon dispersion
relation (which serves for determining the renormalisation of the bare (input)
anisotropy), of the static quark-antiquark potential, of the deconfining phase
transition and finally of the spectrum of low-lying glueballs in pure gauge
theory.

It turns out that the construction of anisotropic classically perfect
gauge actions is feasible. Measuring the renormalised
anisotropy using the torelon dispersion relation turns out to be
stable and unambiguous and shows that the renormalisation of the
anisotropy is small and under good control. The measurements of the
static quark-antiquark potential indicate that the violations of
rotational symmetry are small if the (spatial) lattice is not
exceptionally coarse. This shows that the parametrisation describes
accurately the true classically perfect action, which is known to have
good properties concerning rotational symmetry \cite{Blatter:1996ti}.
The study of the glueball spectrum is facilitated a lot due to the
anisotropic nature of the action, even for (rather small) anisotropy
$\xi=2$. Results, including continuum extrapolations, are obtained for
glueball states having much larger mass than the highest-lying states
that could have been resolved with the same amount of computational
work using the isotropic FP action.  However, with the statistics
reached so far, it is not yet possible to conclude whether the scaling
properties of the glueball states, i.e.~the behaviour of the measured
energies as the lattice spacing is changed, are definitely better
for the classically perfect action as compared to the mean-link and
Symanzik improved anisotropic action
\cite{Morningstar:1997ff,Morningstar:1999rf}. Especially, this is the
case for the lowest-lying scalar glueball where the presence of a
critical end point of a line of phase transitions in the
fundamental-adjoint coupling plane, near the fundamental axis, causes
large distortions of the scaling behaviour (sometimes called the
``scalar dip'').  Our action includes in its rich structure operators
transforming according to the adjoint representation.  If their
coupling (which we do not control specifically during the construction
and parametrisation) lies in a certain region, the effect of the
critical end-point on scalar quantities at certain lattice spacings
may even be enhanced compared to other (more standard) discretisations
with purely fundamental operators.  Concerning the lattice artifacts
observed in the glueball simulations, we have to add that other
effects, e.g.~due to the finite size of the lattices, may also be
present and that this issue requires further study. Being the first
application of the FP action technique to anisotropic lattices,
however, it goes beyond the scope of this work to systematically check
all the possible sources of errors in the glueball mass
determinations.  Furthermore, we note that the computational overhead of the
parametrised anisotropic classically perfect action compared to the
standard Wilson action as well as to the mean-link and Symanzik
improved action is considerable.

The paper is organised as follows. In section \ref{ch:construction} we
describe our method of generating anisotropic perfect gauge actions without
having to perform a certain number of renormalisation group transformations
(RGT) leading from very fine to coarse lattices typical for MC simulations.
Instead, we make use of the parametrisation of the isotropic FP action
presented in \cite{Niedermayer:2000yx,Wenger:2000aa,Niedermayer:2000ts}.  We
start with the \emph{isotropic} parametrised FP action on rather large
fluctuations and perform a small number of purely \emph{spatial} blocking
steps. If a scale 2 block transformation is used, the resulting actions have
anisotropies $\xi=$~2, 4, 8, $\ldots$. The parametrisation ansatz used for
these actions is a generalisation of the isotropic parametrisation and is
described as well.

Section \ref{ch:xi2act} contains the main body of the present work, that is,
the explicit construction and the application of a classically perfect gauge
action on $\xi=2$ anisotropic lattices. The results from measurements of
physical quantities are presented there: the anisotropy of the 
action is measured using the torelon dispersion relation, the (spatial) scale
is determined by the static quark-antiquark potential, where for some values
of the coupling $\beta$, off-axis separations of the quarks are considered as
well, in order to estimate violations of rotational symmetry by the
parametrised action. Another way of determining the (temporal) scale is the
study of the deconfining phase transition, measuring the critical couplings
$\beta_{\text{crit}}$ for different temporal extensions $N_t$ of the lattice.
Finally, the glueball spectrum is determined by performing simulations at
three different lattice spacings using the $\xi=2$ action and taking the
continuum limits whenever it is possible.

The feasibility of iterating the procedure, i.e.~adding one more spatial
blocking step which leads to a classically perfect action with anisotropy
$\xi=4$, is briefly checked in section \ref{ch:xi4act}. It contains the
explicit construction and parametrisation of the $\xi=4$ action and the
determination of the renormalised anisotropy via the torelon dispersion
relation.

Section \ref{ch:Conclusions} finally contains a summary of the main results,
some conclusions and an outlook to future work.

The parameters of the classically perfect
actions used throughout this work are collected in appendix \ref{app:actions},
while appendix \ref{app:rescoll} contains the detailed results from our simulations.

\section{The Construction of Perfect Anisotropic Actions}
\label{ch:construction}

\subsection{Introduction}
The main idea for the construction of (effective) actions on
anisotropic lattices is to start from an isotropic lattice and to
perform one or more purely spatial renormalisation group
transformations (RGT). Each scale two RGT step doubles the
spatial lattice spacing but preserves the spacing in temporal
direction. We are therefore naturally lead from an action on an
isotropic lattice to (effective) actions on anisotropic lattices with
anisotropies $\xi =2,4,8,\ldots$. Of course, choosing the scale $s$ of
the RGT different from two, we can in principle reach any desired
anisotropy $\xi=s,s^2,s^3,\ldots$.

A qualitative picture of the construction of classically perfect
actions on anisotropic lattices is the following. We start from an
isotropic FP action defined as a point in the infinite dimensional
space of all couplings on the critical surface where the correlation
length is infinite ($\beta \rightarrow \infty$). By performing one (or
more) spatial RGT we move away from that point always staying on the
critical surface. Since the spatial blocking kernel is zero on all
classical configurations, the classical properties of the action are
preserved exactly, i.e.~it has no lattice artifacts on the solutions
of the (anisotropic) lattice equations of motion.  (One can say that
the action is an on-shell tree-level Symanzik improved action to all
orders in $a$.)  It is, however, important to note that the resulting
anisotropic classically perfect (AICP) action is, generally speaking,
no longer a FP action, i.e.~the fixed point of a renormalization group
transformation. However, it still possesses all the good properties of
a classically perfect action, in particular it preserves scale
invariant solutions of the classical equations of motion,
i.e.~classical instanton solutions. Moreover, since the spatial RGT
steps are performed on the critical surface at $\beta \rightarrow
\infty$, the transformation reduces to a saddle point problem
representing an implicit equation for the anisotropic classically
perfect action similar to the FP equation in the isotropic case.

In order to check whether this qualitative picture is correct and that
the construction works in practice, we have analytically calculated
the AICP actions for $d=2$ scalar field theory (for earlier studies on
this subject, see \cite{Bietenholz:1999kr}) as well as for the
quadratic approximation to $d=4$ gauge theory for generic anisotropies
$\xi$. In these two cases it turns out that starting from an
anisotropic action for very smooth configurations and performing
isotropic RGTs or starting from an isotropic FP action and performing
an anisotropic RGT at the end is equivalent and leads to the same
classically perfect anisotropic action. It also turns out that the
quality of the results are as good as what has been obtained with the
isotropic FP action \cite{DeGrand:1995ji,Blatter:1996ti}. In
particular we do not see any deviation from the classical continuum
dispersion relation and almost no violation of the rotational symmetry
in the perturbative potential for $r/a > 1$. For further details we
refer to \cite{Rufenacht:2001aa}.

In practice the main advantage of applying one (or more) anisotropic spatial
RGT is that we can directly start from a recent parametrisation of the
isotropic FP gauge action
\cite{Niedermayer:2000yx,Wenger:2000aa,Niedermayer:2000ts} which is valid on
rather coarse lattices. We can therefore avoid the cumbersome cascade
procedure leading from very fine to coarser and coarser lattices.

In the following we first describe the anisotropic spatial blocking in section
\ref{subsec:spatialblocking} and then present an extension of the isotropic
parametrisation to include anisotropy in section \ref{sec:par}.

\subsection{The Spatial Blocking}\label{subsec:spatialblocking}

The isotropic FP action is defined by the saddle point equation (at $\beta=\infty$):
\begin{equation}
  \label{eq:FP_equation}
  {\mathcal A}^{\text{FP}}(V) = \min_{\{U\}} \left\{{\mathcal A}^{\text{FP}}(U) + T(U,V)\right\},
\end{equation}
where $T(U,V)$ is the blocking kernel connecting the fine gauge configurations $U$
to the coarse configurations $V$:
\begin{equation}\label{eq:isotropic blockin kernel}
T(U,V)=\sum_{n_B,\mu}\left
( \mathcal{N}_\mu^\infty(n_B)-\frac{\kappa}{N}\text{Re Tr}
[V_\mu(n_B)Q_\mu^\dag(n_B; U)]\right) ,
\end{equation}
where the normalisation term $\mathcal{N}_\mu^\infty(n_B)$ is given by
\begin{equation}\label{eq:rgtnorm}
\mathcal{N}_\mu^\infty(n_B)=\max_{W\in \text{SU(N)}}\left\{\frac{\kappa}{N}\text{Re Tr}[WQ_\mu^\dag(n_B; U)]\right\},
\end{equation}
while $Q_\mu(n_B; U)$ is the blocked link,
\begin{equation}
Q_\mu(n_B; U)=W_\mu(2n_B; U)W_\mu(2n_B+\hat{\mu}; U).
\end{equation}
Here $W_\mu(n; U)$ denotes a smeared (fuzzy) fine link, which is constructed
as a sum of simple staples as well as of diagonal staples along the planar
and spatial diagonal directions orthogonal to the link (RGT 3 transformation)
\cite{Blatter:1996ti}, respectively.

To perform the blocking only spatially, doubling the lattice spacing
in spatial direction while leaving the temporal lattice spacing
unchanged, we use a purely spatial blocking kernel
$T_{\text{sp}}(U,V)$ obtained by setting in the isotropic definition
\begin{equation}
 Q_4(n_B; U)=W_4(2n_B; U)
\end{equation}
thus doing only a smearing and no blocking in temporal direction. (In
the case of spatial blocking the expression ``$2n_B$'' stands for
$(2n_B^1,2n_B^2,2n_B^3,n_B^4)$.)  In addition, we have the freedom to
choose different values of $\kappa$ for the spatial ($\kappa_s$) and
temporal ($\kappa_t$) blocking.

\subsection{The Parametrisation} \label{sec:par}

The classically perfect actions, described in principle by an
infinite number of couplings, have to be parametrised in order to be
suitable for numerical simulations.  We have shown previously
\cite{Niedermayer:2000yx,Wenger:2000aa,Niedermayer:2000ts} that a
parametrisation of the isotropic FP gauge action including APE-like
smearing behaves much better and is much more flexible compared to
common parametrisations using traces of closed loops with comparable
computational cost. We thus decide to use an extension of this parametrisation
to describe the anisotropic classically perfect gauge action.

For convenience, let us quickly review the parametrisation of the
isotropic action. The parametrisation uses mixed polynomials of traces
of simple loops (plaquettes) built from single gauge links $U_\mu(n)$
as well as from asymmetrically (APE-like) smeared links
$W_\mu^{(\nu)}(n)$. We introduce the notation $S_\mu^{(\nu)}(n)$ for the
sum of two staples of gauge links connecting two lattice sites in
direction $\mu$ lying in the $\mu\nu$-plane:
\begin{equation}
S_\mu^{(\nu)}(n)=U_\nu(n)U_\mu(n+\hat\nu)U_\nu^\dagger(n+\hat\mu)+U_\nu^\dagger(n-\hat\nu)U_\mu(n-\hat\nu)U_\nu(n-\hat\nu+\hat\mu).
\end{equation}

To build a plaquette in a plane $\mu\nu$ from smeared links it is
convenient to introduce \emph{asymmetrically} smeared links. First
define\footnote{The argument $n$ is suppressed in the following.}
\begin{equation}
\label{eq:asym1}
Q_\mu^{(\nu)}=\frac{1}{4}\left(\sum_{\lambda\ne \mu,\nu} S_\mu^{(\lambda)}+\eta S_\mu^{(\nu)}\right)
- \left(1+\frac{1}{2}\eta\right)U_\mu.
\end{equation}
Out of these sums of matrices connecting two neighboring points $n$, $n+\hat\mu$, we build the
asymmetrically smeared links 
\begin{equation}
\label{eq:smlconstr}
W_\mu^{(\nu)}=U_\mu+ \sum_{i=1} c_i Q_\mu^{(\nu)} \large(U_\mu^\dagger
Q_\mu^{(\nu)}\large)^{i-1},
\end{equation} 

The parameters $\eta$, $c_i$ used for the smearing may depend on local fluctuations measured by
$x_\mu(n)$ defined as
\begin{equation}
x_\mu(n)=\text{Re Tr}(Q_\mu^{\text{s}}(n)U_\mu^\dagger(n)),
\end{equation}
with the symmetrically smeared link
\begin{equation}
Q_\mu^{\text{s}}(n)=\frac{1}{6}\sum_{\lambda\ne\mu}S_\mu^{(\lambda)}(n)-U_\mu(n).
\end{equation}
This parameter is negative, $-4.5\leq x_\mu(n) \leq 0$, and it vanishes
for trivial gauge configurations.

The smearing parameters are chosen to be polynomials of $x_\mu$
with free coefficients (determined later by a fit to the FP action):
\begin{eqnarray}\label{eq:eta_x}
\eta & = & \eta^{(0)}+\eta^{(1)}x+\eta^{(2)}x^2+\cdots,\\
c_i & = & c_i^{(0)}+c_i^{(1)}x+c_i^{(2)}x^2+\cdots.\label{eq:c_x}
\end{eqnarray}

Of course, the asymmetrically smeared links $W_\mu^{(\nu)}$ built out of a large
number of paths connecting the neighbouring lattice sites are no
longer elements of the SU(3) gauge group. They might be projected back
to SU(3), however this task increases the computational cost in actual
numerical simulations. Moreover, our studies have shown that
projection reduces the degrees of freedom in defining the action and we
are thus using the smeared links $W_\mu^{(\nu)}$ as they are.

We now build a smeared plaquette variable,
\begin{equation}
w_{\mu\nu}=\text{Re Tr}(1-W_{\mu\nu}^{\text{pl}}),
\end{equation}
as well as the ordinary one,
\begin{equation}
u_{\mu\nu}=\text{Re Tr}(1-U_{\mu\nu}^{\text{pl}}),
\end{equation}
where
\begin{equation}
W_{\mu\nu}^{\text{pl}}(n)=W_\mu^{(\nu)}(n)W_\nu^{(\mu)}(n+\hat\mu)W_\mu^{(\nu)\dagger}(n+\hat\nu)W_\nu^{(\mu)\dagger}(n)
\end{equation}
and
\begin{equation}
U_{\mu\nu}^{\text{pl}}(n)=U_\mu(n)U_\nu(n+\hat\mu)U_\mu^\dagger(n+\hat\nu)U_\nu^\dagger(n),
\end{equation}
respectively.
Finally, the action is built from these plaquette variables using a
mixed polynomial ansatz of the form
\begin{equation}
\label{eq:isoacpol}
\mathcal{A}[U]=\frac{1}{N_c}\sum_n\sum_{\mu<\nu}\sum_{k,l}p_{kl}u_{\mu\nu}(n)^k
w_{\mu\nu}(n)^l,
\end{equation}
where the coefficients $p_{kl}$ are again free parameters defined by
a fit to the FP action.

\begin{figure}[h]
\begin{center}
\begin{tabular*}{\textwidth}[c]{c@{\extracolsep{\fill}}ccc}
\includegraphics[width=2.25cm]{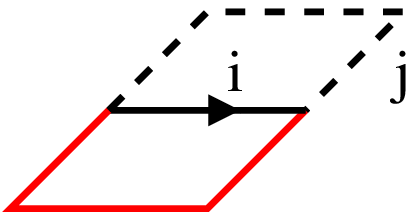} &
\includegraphics[width=2.25cm]{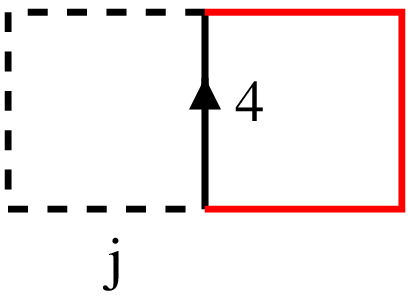} &
\includegraphics[width=1.6875cm]{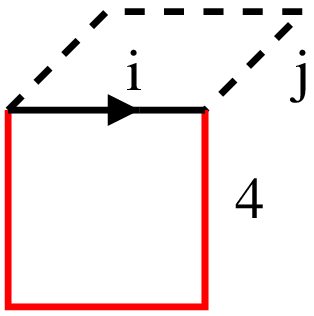} &
\includegraphics[width=1.35cm]{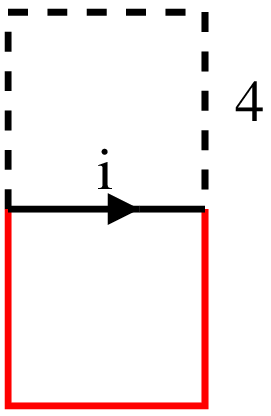}\\
$\eta_1$ & $\eta_2$ & $\eta_3$ & $\eta_4$\\
\end{tabular*}
\end{center}
\caption{Asymmetry in the construction of the smeared link matrices
$Q_\mu^{(\nu)}$. \emph{From left to right}: spatial staple
contributing to a smeared spatial link in a spatial plaquette
($\eta_1$), temporal staple contributing to a smeared temporal link
($\eta_2$), temporal staple contributing to a smeared spatial link in
a spatial plaquette ($\eta_3$), temporal staple contributing to a
smeared spatial link ($\eta_4$).}
\label{fig:etas}
\end{figure}

In order to adapt this parametrisation of the isotropic FP action to 
anisotropic actions, we use three
(rather straightforward) extensions.  Firstly, the coefficients
$p_{kl}$ in eq.~(\ref{eq:isoacpol}) are chosen differently depending
on the orientation of the plaquette $\mu\nu$, i.e.,
$p_{kl}^{\text{sp}}$ for $\mu\nu \in \{12, 13, 23\}$ (spatial
plaquettes) and $p_{kl}^{\text{tm}}$ for $\mu\nu \in \{14, 24, 34\}$
(temporal plaquettes).  Secondly, the parameter $\eta$ entering in
eq.~(\ref{eq:asym1}) describing the asymmetry between differently oriented
staples contributing to a smeared link $W_\mu^{(\nu)}$ is generalised to
distinguish between smeared spatial and
temporal links:
\begin{eqnarray}
Q_i^{(j)} & = & \frac{1}{4}(\sum_{k\ne i,j} S_i^{(k)}+\eta_1 S_i^{(j)}+\eta_3 S_i^{(4)})-\frac{1}{2}(1
+\eta_1+\eta_3)U_i,\\
Q_i^{(4)} & = & \frac{1}{4}(\sum_{\lambda\ne i,4} S_i^{(\lambda)}+\eta_4 S_i^{(4)})-(1
+\frac{1}{2}\eta_4)U_i,\\
Q_4^{(j)} & = & \frac{1}{4}(\sum_{\lambda\ne 4,j} S_4^{(\lambda)}+\eta_2 S_4^{(j)})-(1
+\frac{1}{2}\eta_2)U_4,
\end{eqnarray}
where $i,j,k=1,2,3$ and $\mu,\nu,\lambda=1,\ldots,4$.
The anisotropic parameters $\eta_1,\ldots,\eta_4$ may be again
polynomials in the local fluctuation parameter $x_\mu$. These situations
are depicted in Figure \ref{fig:etas}.

\begin{figure}[h]
\begin{center}
\begin{tabular*}{\textwidth}[c]{c@{\extracolsep{\fill}}cc}
\includegraphics[width=1.6875cm]{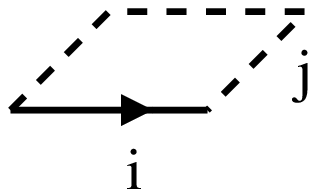}\hspace{1cm} &
\includegraphics[width=1.35cm]{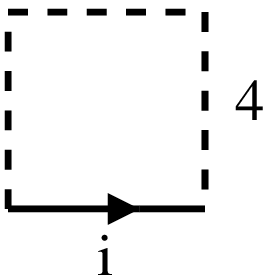}\hspace{1cm} &
\includegraphics[width=1.35cm]{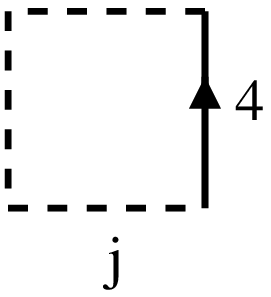}\\
$c_{i1}$ & $c_{i2}$ & $c_{i3}$\\
\end{tabular*}
\end{center}
\caption{Asymmetry in the construction of the smeared links
$W_\mu^{(\nu)}$ from the matrices $Q_\mu^{(\nu)}$. \emph{From left to
right}: spatial $Q_i^{(j)}$ contributing to a smeared spatial
plaquette ($c_{i1}$), spatial $Q_i^{(4)}$ contributing to a smeared temporal plaquette ($c_{i2}$), temporal $Q_4^{(j)}$ ($c_{i3}$).}
\label{fig:cis}
\end{figure}

Finally, the construction of the smeared links $W_\mu^{(\nu)}$
from the matrices $Q_\mu^{(\nu)}$, described by the parameters $c_i$
in eq.~(\ref{eq:smlconstr}), is generalised such that these parameters
are chosen differently for the construction of temporal links (always
contributing to temporal
plaquettes), spatial links contributing to spatial plaquettes and
spatial links contributing to temporal plaquettes, $c_{i1}$, $c_{i2}$
and $c_{i3}$, respectively, see Figure \ref{fig:cis}.

It is not a priori clear whether all the three extensions are
necessary at the same time. Results from fits to different extended
parametrisations (Table \ref{tab:fitqual}), however, show that it is
indispensable to have the full parameter set. The values of $\chi^2$
of the fit to the true classically perfect actions as well as the
measured renormalised anisotropies of the parametrised action (see
Section \ref{sec:xir}) indicate that the parametrisation is not
flexible enough neither for $\eta_3=\eta_4\equiv 0$ (leading to a
positive definite transfer matrix connecting only neighbouring time
slices) nor if the parameters $c_{ij}$,
$j=1,2,3$ are set to be equal.

\begin{table}[htbp]
      \begin{center}
    \renewcommand{\arraystretch}{1.3}
    \begin{tabular*}{\textwidth}[c]{r@{\extracolsep{\fill}}rrccrr}
\hline
\# $\eta$ & \# $c_i$ & $n$ & $\max(k+l)_{\text{sp}}$ & $\max(k+l)_{\text{tm}}$ & $\chi_{\text{d}}^2$ & $\xi_{\text{R}}$\\
\hline
4 & 1 & 4 & 4 & 4 & 0.0250 & 1.63(2)\\
2 & 3 & 3 & 4 & 4 & 0.0238 & \\
4 & 3 & 3 & 4 & 4 & 0.0144 & 1.912(9)\\
\hline
\end{tabular*}
\caption{Comparison of the accuracy $\chi_{\text{d}}^2$ of the fit on
derivatives on 20 configurations at $\beta=3.5$ and the measured
renormalised anisotropy $\xi_R$ (using the torelon dispersion
relation) at $\beta=3.3$ for different choices of the set of
non-linear parameters.}  \end{center}
\label{tab:fitqual}
\end{table}

Our anisotropic action shall have the correct normalisation and
anisotropy in the continuum limit.  This can be accomplished by demanding
the following two normalisation conditions to be exactly fulfilled \cite{Rufenacht:2001aa}:

\begin{eqnarray}
p^{\text{st}}_{01}+p^{\text{st}}_{10}+2p^{\text{st}}_{01}c_{13}+p^{\text{st}}_{01}c_{13}\eta_2+
2p^{\text{ss}}_{01}c_{11}\eta_3+p^{\text{st}}_{01}c_{12}\eta_4 & = & \xi,\\
p^{\text{ss}}_{01}+p^{\text{ss}}_{10}+2p^{\text{ss}}_{01}c_{11}+
2p^{\text{st}}_{01}c_{12}+2p^{\text{ss}}_{01}c_{11}\eta_1 & = & \frac{1}{\xi},
\end{eqnarray}
where all parameters $\eta$, $c$ denote the constant ($0^{\text{th}}$
order) terms in the polynomials in $x_\mu$.

Let us note that the parametrisation presented here leads to an overhead in
the updates in MC simulations of 66 compared to the Wilson action and
of 12 compared to the tadpole and tree-level improved action.

\section{The $\mathbf{\xi=}2$ Perfect Action} \label{ch:xi2act}

\subsection{Construction} \label{sec:xi2constr}

We construct the $\xi=2$ classically perfect anisotropic action
starting from the parametrised isotropic action
\cite{Niedermayer:2000yx} performing one spatial blocking step as
described in Section \ref{ch:construction}. Studies of the quadratic
approximation suggest $\kappa_t=\xi^2 \kappa_s=4\kappa_s$ in the RG
transformation for keeping the block transformation as close to the
isotropic case as possible. Furthermore, spatial locality should stay
at its optimum for the ``isotropic'' value of $\kappa_s=8.8$
\cite{Blatter:1996ti}, thus we choose $\kappa_t=35.2$.  Indeed, it
turns out that varying $\kappa_t$ away from $\kappa_t=35.2$, keeping
$\kappa_s=8.8$ fixed, makes it much more difficult to parametrise the
resulting perfect action such that the renormalised anisotropy
$\xi_{\text{R}}$ stays close to the input value $\xi=2$.  Coarse
configurations $V$ are generated using the ad-hoc anisotropic action
(see Appendix \ref{app:adhoc}), the FP eq.~(\ref{eq:FP_equation})
including the purely spatial blocking kernel $T_{\text{sp}}(U,V)$ is
used as a recursion relation with the intermediate isotropic action
(see Appendix \ref{app:isointact}) on the r.h.s.~yielding the
classically perfect $\xi=2$ action on the l.h.s.  Examining the
blocking, we observe that coarse configurations obtained in a MC run
with the ad-hoc anisotropic action are mapped to fine configurations
that are close to isotropic (concerning spatial and temporal plaquette
values and expectation values of the Landau gauge-fixed link
variables) for input anisotropy $\xi_{\text{ad-hoc}}\approx$~3.2.  We
thus generate 20 configurations each at $\beta_{\text{ad-hoc}}=$~2.5,
3.0, 3.5, 4.0 using the ad-hoc anisotropic action\footnote{Note, that
the values of $\beta_{\text{ad-hoc}}$, used in this section, may not
be directly compared to the values of $\beta_{\text{perf}}$
corresponding to the parametrised classically perfect action, used in
the following sections; but rather the fluctuations (plaquette values,
expectation values of Landau gauge-fixed links etc.) of the gauge
configurations produced in MC runs using the perfect action should lie
approximately in the same range as the fluctuations of the initial
coarse configurations produced with the ad-hoc action.} with
$\xi_{\text{ad-hoc}}$=3.2. These configurations are spatially blocked
using the intermediate isotropic action (see Appendix \ref{app:isointact})
which describes the minimised isotropic configurations on the
r.h.s. of eq.~(\ref{eq:FP_equation}) reasonably well.

To construct the action, we perform several non-linear fits for the
derivatives of 20 configurations at $\beta_{\text{ad-hoc}}=3.5$, using different sets
of parameters. The parameters $c_i$ are chosen to be non-zero for
$i=1,2,3$ as adding additional free $c_4$ parameters does not improve
$\chi^2$ significantly.

Having fixed the non-linear parameters $\eta$ and $c_i$, we include
the derivatives of 20 configurations at $\beta_{\text{ad-hoc}}=4.0$ and the action
values of all the configurations at $\beta_{\text{ad-hoc}}=$~3.5, 4.0 in the linear
fit of the parameters $p_{kl}$, keeping the non-linear parameters
fixed. The resulting values of $\chi^2$ and the linear behaviour of
the actions suggest a linear set with $\max(k+l)_{\text{sp}}=4$,
$\max(k+l)_{\text{tm}}=3$ where the action values are included with
weights $w_{\text{act}}=0.018$ relative to the derivatives.  Using
this value, the good parametrisation of the derivatives obtained in
the full nonlinear fit is preserved, at the same time the mean error
of the action value due to the parametrisation is as small as 0.35\%.
It is checked that the number of data points in the fit is large enough,
making sure that the values of $\chi^2$ as well as the errors of the
action values do not increase significantly on configurations independent
of those used in the fit. Furthermore, these quantities are determined on
configurations down to $\beta_{\text{ad-hoc}}=3.0$ and it turns out that the action works
down to this value of $\beta_{\text{ad-hoc}}$.
Finally, it is checked that the linear parameters
$p_{kl}$ do not lead to artificial gauge configurations in MC
simulations caused by (local) negative contributions to the total
action.  The parameters of the resulting action are listed in Appendix
\ref{app:xi2act}.

\subsection{The Renormalised Anisotropy} \label{sec:xir}\label{sec:smearing}
The renormalised anisotropy $\xi_R$ of the action presented above is
measured using the torelon dispersion relation as described in
\cite{Teper:1998te,Alford:2000an}. The torelon is a closed gluon flux
tube encircling the periodic spatial boundary of size $L$ of a
torus. It is created and annihilated by closed gauge operators winding
around this boundary.  String models which are a good approximation
for long flux tubes predict its energy to behave as
\begin{equation} \label{eq:stringtorelon}
E(L)=\sigma L + \frac{\pi}{3L},
\end{equation}
where $\sigma$ is the string tension and the second term describes the
string fluctuation for a bosonic string
\cite{Luscher:1981ac,Isgur:1985bm,Perantonis:1988wj,Perantonis:1989uz}.
The string models assume no restriction on the transverse modes of the
string, thus they apply to volumes with large transverse extension $S$.
Due to this reason and because we prefer not to have too large transverse
momenta, we work on $S^2\times L\times T$ lattices, where $S$ and $T$ are
large, whereas the length of the $z$-direction $L$ is chosen to be rather
short such that the torelon energy (see eq.~(\ref{eq:stringtorelon})) is
not too heavy.
Since the colour flux tube has finite width in physical units, it is useful to
employ (iteratively)
APE smeared links \cite{Albanese:1987ds} to improve
the overlap of the operators with the ground state.  One step of APE
smearing acts on a spatial gauge link $U_j(x)$ as 
\begin{eqnarray}
\label{eq:APE}
  {\mathcal S}_1 U_j(x) & \equiv & {\mathcal P}_{\text{SU(3)}} 
   \Big\{ U_j(x) + \lambda_s \sum_{k \neq j}
  (U_k(x) U_j(x+\hat k) U_k^\dagger(x+\hat j)  \\
 & &  \hspace{2cm}  + U_k^\dagger(x-\hat k)
  U_j(x-\hat k) U_k(x-\hat k+\hat j))  \Big\}, \nonumber 
\end{eqnarray}
the original link variable is replaced by itself plus a sum of the
four neighbouring spatial staples and then projected back to SU(3). The
smeared and projected links ${\mathcal S}_1U_j(x)$ have the same symmetry
properties under gauge transformations, charge conjugation and
reflections and permutations of the coordinate axes as the original gauge
links. The smearing is
used iteratively allowing to measure operators on 
different smearing levels, ${\mathcal S}_n U$.

We measure the operators
\begin{equation}
T_n(x,y,t)=\text{Tr} \prod_{z=0}^{L-1}{\mathcal S}_n U_z(x,y,z,t),
\end{equation}
for all $x$, $y$, $t$, where ${\mathcal S}_n U_z$ is the $n$ times
iteratively smeared link in the longitudinal $z$ direction. By
discrete Fourier transform, we project out the state with momentum
$\mathbf{p}=(p_x,p_y)=(n_x,n_y)(2\pi/Sa_s)$:
\begin{equation}
T_n(\mathbf{p},t)=\sum_{x,y}T_n(x,y,t)e^{i(p_xx+p_yy)}
\end{equation}
and build the correlators
\begin{equation}
C_{nn'}(\mathbf{p},t)=\frac{1}{N_t}\sum_{\tau}\langle T_n(\mathbf{p},\tau) T_{n'}(-\mathbf{p},\tau+t)\rangle.
\end{equation}
From these correlators of operators measured on the different smearing levels we obtain the
energy values $a_t E(\mathbf{p})$ using variational methods (see Appendix B of \cite{Niedermayer:2000yx}).
The continuum dispersion relation 
\begin{equation}
E^2=\mathbf{p}^2+m^2
\end{equation}
on the lattice becomes (in temporal units)
\begin{equation}
(a_t E)^2 = a_t^2(\mathbf{p}^2+m^2)=\frac{(a_s\mathbf{p})^2}{\xi_{\text{R}}^2}+(ma_t)^2=\frac{1}{\xi_{\text{R}}^2}(n_x^2+n_y^2)\left(\frac{2\pi}{S}\right)^2
+(ma_t)^2,
\end{equation}
where $n_x$, $n_y$ are the components of the (transversal) lattice
momentum. On an anisotropic lattice, this equation allows for the
extraction of the renormalised anisotropy $\xi_{\text{R}}=a_s/a_t$
(measuring the ``renormalisation of the speed of light'') as well as
the torelon mass $ma_t$, which in turn may be used to get an estimate
of the scale using eq.~(\ref{eq:stringtorelon}) and known values of
the string tension $\sigma$.  The main advantage of this approach is
obtaining the renormalised anisotropy $\xi_R$ as well as an estimate
for the scales $a_s$ and $a_t$ by performing computationally rather
inexpensive measurements.

The parameters of the simulations carried out are collected in Table
\ref{tab:xi2tordet}.

\begin{table}[tbhp]
\renewcommand{\arraystretch}{1.3}

  \begin{center}
    \begin{tabular*}{\textwidth}[c]{c@{\extracolsep{\fill}}cc}
      \hline\vspace{-0.05cm} 
      $\beta$ & $S^2\times L\times T$ & \# sweeps / measurements\\
      \hline
      3.00     & $8^2\times 4\times 20$ & 72000 / 14400\\
      3.15$^a$ & $8^2\times 5\times 20$ & 54000 / 10800\\
      3.15$^b$ & $5^2\times 8\times 20$ & 47600 / 9520\\
      3.30     & $12^2\times 6\times 30$ & 35600 / 7120\\
      3.50$^a$ & $12^2\times 8\times 24$ & 39400 / 7880\\
      3.50$^b$ & $14^2\times 6\times 30$ & 36800 / 7360\\
      \hline
    \end{tabular*}
    \caption{Run parameters for the torelon measurements using the $\xi=2$ perfect
      action. The lattice extensions in torelon direction $L$, the extension
      in the two transversal spatial directions $S$ as well as the temporal extension
      $L$ are given.}
    \label{tab:xi2tordet}
  \end{center}
\end{table}

The Polyakov line around the short spatial direction is measured
using operators at
smearing levels ${\mathcal S}_n U$ with $n = 3, 6, 9, 12, 15$ for
$\beta\ge 3.3$ and $n = 2, 4, 6, 8, 10$ for $\beta\le 3.15$,
respectively, and smearing parameter
$\lambda_s=0.1$.  The measured energies $E(p^2)$  are given,
together with the number of operators used in the variational method,
the fit ranges and the values of $\chi^2$ per degree of freedom,
$\chi^2/N_{\text{DF}}$, in Tables \ref{tab:xi2torcoll} and 
\ref{tab:xi2torcoll2} in Appendix \ref{app:rescoll}.

Figure \ref{fig:tordisp_b33_018} displays an example of a dispersion
relation at $\beta=3.3$ including all values of $p$ at which the
energies could be determined on the given lattice. Note, that the
energies of the $p=0$ torelons (the torelon masses) may be hard to
determine because the effective masses do not reach a plateau within
the temporal extent of the lattice, probably due to the small overlap of the
operators used with these states; however, using the non-zero
momentum energies, the masses may still be accurately determined.

To determine the renormalised anisotropy $\xi_R$ as well as the
torelon mass in units of the temporal lattice spacing $m a_t$,
we perform fits to the dispersion relation, taking into account
the correlations between different operators. The range of $p^2$
considered is chosen depending on $\chi^2/N_{\text{DF}}$ of the fit
and the precision of the dispersion relation data in the respective
range. The results are given in Table \ref{tab:xi2torres}.

\begin{table}[htbp]
\renewcommand{\arraystretch}{1.3}
  \begin{center}
    \begin{tabular*}{\textwidth}[c]{c@{\extracolsep{\fill}}cccccccc}
      \hline\vspace{-0.05cm}
      $\beta$ & fit range & $\xi_R$ & $m_{\text{T}} a_t$ & $\chi^2/N_{\text{DF}}$\\
      \hline
      3.00     & 0..5 & 1.903(81) & 1.324(98) & 0.39\\
      3.15$^a$ & 0..4 & 1.966(39) & 0.700(35) & 0.83\\
      3.15$^b$ & 0..5 & 2.022(104) & 1.262(84) & 0.62\\
      3.30     & 1..9 & 1.912(9) & 0.311(5) & 0.94\\
      3.50$^a$ & 1..10 & 1.836(9) & 0.149(10) & 0.94\\
      3.50$^b$ & 1..8 & 1.826(16) & 0.208(16) & 1.61\\
      \hline
    \end{tabular*}
    \caption{Results of the torelon simulations using the $\xi=2$ perfect action. The fit
      range in $p^2$ is given in units of $(2\pi/S)^2$.}
    \label{tab:xi2torres}
  \end{center}
\end{table}

Using the finite size relation for the torelon mass corresponding to
eq.~(\ref{eq:stringtorelon}),
\begin{equation}\label{eq:finite_size_torelon}
M_T(La_s)=(\sigma+\frac{D}{(La_s)^2})La_s \, ,
\end{equation}
we calculate the string tension $\sigma$ and thus the spatial scale
$a_s$.  The string model predicts $D=-\frac{\pi}{3}$ for long strings
thence we use this string picture value and stay aware that the
estimate gets worse for short strings. To obtain the hadronic scale
$r_0$ and the lattice spacings we employ $r_0\sqrt{\sigma}$=1.193(10)
from \cite{Niedermayer:2000yx} and use the definition
$r_0=0.50$~fm. The results are collected in Table
\ref{tab:xi2torscal}.

The two simulations denoted by $3.15^b$ and $3.50^b$ are used to check the
stability of the method. The first one is carried out on a lattice
with small transversal spatial size (the torelon string measured is
even longer), the latter is measuring the energies of a rather short
string.
The scale is measured accurately using the static quark-antiquark
potential (see Section \ref{sec:xi2scale}) so that we can compare our
torelon estimates to the much more reliable values in
Table \ref{tab:xi2scalcomp}. It turns out that the
deviation of the torelon estimates for the scale $r_0/a_s$ from the
potential values does not exceed 6\% if the length
of the string is not too short (say $La_s\gtrsim 1$~fm). For short
strings, the string picture does not apply and we may therefore not
expect good results in this case. But even for very long strings, it
is not possible to obtain very accurate information about the
scale as long strings are very heavy and thus
difficult to measure (comparable to the long range region of the
static quark-antiquark potential), especially when the anisotropy
$\xi$ is not very large.

There seems to be no problem if the transverse volume is rather small
as in the $\beta=3.15^b$ simulation, except of course the large
momenta occurring which make the determination of the energies more
difficult. As well, the estimates of the renormalised anisotropy
$\xi_R$ do not show significant deviations neither for small
transverse volume nor for short strings.

\begin{table}[htbp]
\renewcommand{\arraystretch}{1.3}
  \begin{center}
    \begin{tabular*}{\textwidth}[c]{c@{\extracolsep{\fill}}cccc}
      \hline\vspace{-0.05cm}
      $\beta$ & $\sqrt{\sigma}a_s$ & $r_0/a_s$ & $a_s$ [fm] & $a_t$ [fm]\\
      \hline
      3.00     & 0.834(20) & 1.43(5)  & 0.350(12) & 0.184(14)\\
      3.15$^a$ & 0.563(8) & 2.12(5)  & 0.236(6) & 0.120(5)\\
      3.15$^b$ & 0.579(18) & 2.06(8) & 0.243(9) & 0.120(11)\\
      3.30     & 0.3578(21) & 3.33(5) & 0.150(2) & 0.079(1)\\
      3.50$^a$ & 0.225(5) & 5.31(16) & 0.094(3) & 0.051(2)\\
      3.50$^b$ & 0.304(7) & 3.92(12) & 0.128(4) & 0.070(3)\\
      \hline
    \end{tabular*}
    \caption{Estimates of the scale determined from torelon results (see Table \ref{tab:xi2torres}),
     see text.}
    \label{tab:xi2torscal}
  \end{center}
\end{table}

\begin{table}[htbp]
\renewcommand{\arraystretch}{1.3}
  \begin{center}
    \begin{tabular*}{\textwidth}[c]{c@{\extracolsep{\fill}}cccc}
      \hline\vspace{-0.05cm}
      $\beta$ & $La_s$ [fm] & $r_0/a_s$ (Torelon) & $r_0/a_s$ (Potential) & rel. error\\
      \hline
      3.00     & 1.48 & 1.43(5) & 1.353(16) & +5.7\%\\
      3.15$^a$ & 1.23 & 2.12(5) & 2.038(2)  & +4.0\%\\
      3.15$^b$ & 1.96 & 2.06(8) & 2.038(2)  & +1.1\%\\
      3.30     & 0.95 & 3.33(5) & 3.154(8)  & +5.6\%\\
      3.50$^a$ & 0.82 & 5.31(16) & 4.906(21) & +8.2\%\\
      3.50$^b$ & 0.61 & 3.92(12) & 4.906(21) & -20.0\%\\
      \hline
    \end{tabular*}
    \caption{The estimates of the scale obtained from torelon measurements (see. Table \ref{tab:xi2torscal}) compared
to the scale measured using the static quark-antiquark potential (see Table \ref{tab:xi2potres}). The length of the
torelon string $La_s$ in physical units is given as well.}
    \label{tab:xi2scalcomp}
  \end{center}
\end{table}

\begin{figure}[htbp]
\begin{center}
\includegraphics[width=10cm]{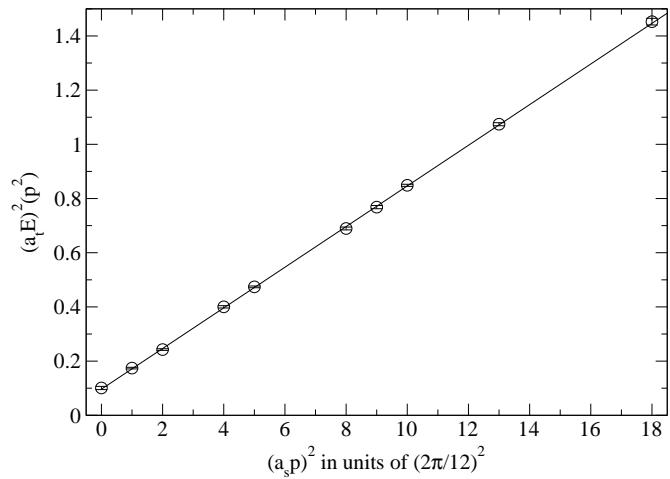}
\end{center}
\caption{Torelon dispersion relation for $\beta=3.3$. The straight line is the correlated fit to $E^2(p)=m_T^2+p^2$ in the range $p^2=1..9$.}
\label{fig:tordisp_b33_018}
\end{figure}


Finally, we may conclude that determining the renormalised anisotropy
using the torelon dispersion relation is a stable and apparently
sensible procedure. There are no manifest problems and it works on
fine lattices as well as on coarse ones. Checks employing different
lattice extensions indicate that the determination of the renormalised
anisotropy is quite stable within a large range of different choices.
The estimate of the lattice scale, however, is not very accurate, mainly
due to the use of the string picture relation,
eq.~(\ref{eq:finite_size_torelon}); the (systematic) error of the
torelon scale is about 5\% for reasonable torelon lengths $L\gtrsim
1$~fm with a tendency of underestimating the lattice spacing $a_s$.

\subsection{The static quark-antiquark potential}
\subsubsection{The Scale}\label{sec:xi2scale}
To set the scale of the action at a given $\beta$-value we measure the
 static quark-antiquark potential, which is reasonably well described by the
phenomenological ansatz \cite{Eichten:1980ms},
\begin{equation}
V(R)=V_0+\frac{\alpha}{R}+\sigma R.
\label{eq:isopot}
\end{equation}
We determine the lattice spacing using the following definition of
$r_c$ \cite{Sommer:1994ce}:
\begin{equation}
r^2 V'(r)|_{r=r_c} = c,
\end{equation}
where the well-known Sommer scale $r_0\approx$~0.50~fm corresponds
to $c=1.65$. For the anisotropic action, we employ the conventional
definition of $r_0$, however, $V(r)$ is now measured in units of 
the temporal lattice spacing and thus we have
\begin{equation}
-\alpha+\xi\hat{\sigma}(\frac{r_c}{a_s})^2=c,
\end{equation}
where $\hat{\sigma}$ denotes the fitted dimensionless value
$\hat{\sigma}=\sigma a_s a_t$ of the string tension. The renormalised
anisotropy $\xi$ thus has to be known before the spatial scale $r_0$
may be determined. For very fine or coarse lattices it may be
advantageous to go to a slightly different
separation corresponding to a different value of the constant
$c$ on the r.h.s. For this purpose, we have collected values
for $c$ and $r_c$ from high-precision measurements of the static
potential performed with the Wilson action
\cite{Bali:1992ab,Edwards:1997xf,Guagnelli:1998ud} listed in Table
\ref{tab:isocrc}.

\begin{table}[htbp]

\renewcommand{\arraystretch}{1.3}
  \begin{center}
    \begin{tabular}{cc}
      \hline\vspace{-0.05cm}
      $r_c/r_0$ & c \\
      \hline
       0.662(1)     & 0.89 \\
       1.00         & 1.65 \\
       1.65(1)      & 4.00 \\
       2.04(2)      & 6.00 \\
      \hline
    \end{tabular}
    \caption{{}Parameter values for the determination of the hadronic
      scale $r_c$ through \mbox{$r^2V'(r)|_{r=r_c} = c$}.}
    \label{tab:isocrc}
  \end{center}
\end{table}

The static quark-antiquark potential is measured on lattices of
different scales $a_s=0.10\ldots 0.37$~fm, employing Wilson loops
built out of APE smeared (see Section \ref{sec:smearing}) spatial
links and simple temporal gauge links.  The measurements at
$\beta=3.00$, 3.30 include the determination of the potential between
quarks that are off-axis separated along the lattice vectors (1,0,0),
(1,1,0), (1,1,1), (2,1,0), (2,1,1), (2,2,1) (and lattice rotations
thereof) in order to estimate violations of rotational symmetry.  The
rest of the measurements include only on-axis separations due to the
large computational cost (concerning speed and memory) of the off-axis
measurement.  The smearing levels used are as follows ($\lambda_f=0.1$
on all the lattices): on the finest lattice at $\beta=3.50$, we use
${\mathcal S}_n U, n = 3, 6, 9, 12, 15$; at $\beta=3.15$ we use
${\mathcal S}_n U,
n = 2, 4, 6, 8, 10$; for the measurements including off-axis
separations, we employ only three different smearing levels each, in
order to save memory and time: ${\mathcal S}_n U, n = 5, 10, 15$ at
$\beta=3.30$ and ${\mathcal S}_n U, n = 2, 4, 6$ on the coarsest lattice at
$\beta=3.00$. The parameters of the simulations are given in Table
\ref{tab:xi2potpar}.

\begin{table}[htbp]
\renewcommand{\arraystretch}{1.3}
  \begin{center}
    \begin{tabular*}{\textwidth}[c]{c@{\extracolsep{\fill}}ccc}
      \hline\vspace{-0.05cm} 
      $\beta$ & off-axis sep. & $S^3\times T$ & \# sweeps / measurements\\
      \hline
      3.00 & yes & $8^3\times 16$ & 42000 / 2800\\
      3.15 & no & $10^3\times 20$ & 39800 / 3980\\
      3.30 & yes & $10^3\times 20$ & 27000 / 1800\\
      3.50 & no & $12^3\times 24$ & 36400 / 3640\\
      \hline
    \end{tabular*}
    \caption{Run parameters for the measurements of the static quark-antiquark potential using the $\xi=2$ perfect
      action.}
    \label{tab:xi2potpar}
  \end{center}
\end{table}

The values of the off-axis potential $a_t V(\vec{r})$ at $\beta=3.30$
are collected in Table \ref{tab:xi2offax} in Appendix
\ref{app:rescoll} and displayed in Figure \ref{fig:pot_grnd}, the
values of the off-axis potential on the coarse lattice at $\beta=3.00$
are collected in Table \ref{tab:xi2offax_b300}. The potential values
$a_t V(r)$ of the on-axis simulations are collected in Table
\ref{tab:xi2onaxpot}.

\begin{figure}[htbp]
\begin{center}
\includegraphics[width=10cm]{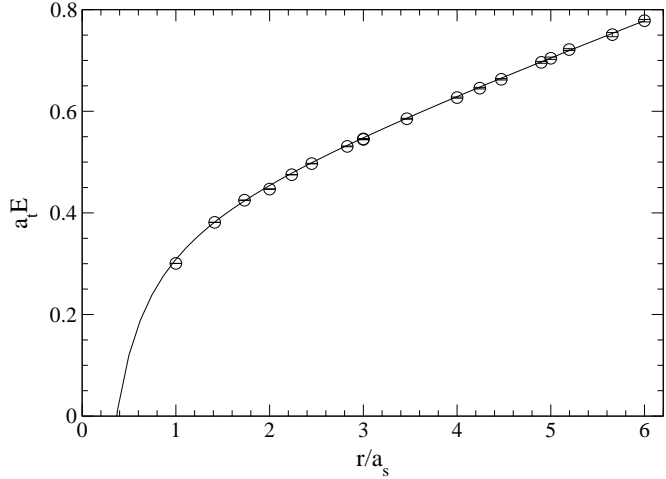}
\end{center}
\caption{Ground state of the static quark-antiquark potential at $\beta=3.3$. Including a fit
to the phenomenological potential ansatz in the range $r=\sqrt{2}..5$.}
\label{fig:pot_grnd}
\end{figure}

The parameters $\alpha$ and $\sigma$ (the string tension) in the
phenomenological potential ansatz, eq.~(\ref{eq:isopot}), are determined using global
fits, the hadronic scale $r_0$ is determined performing local
fits. The fit ranges and results are given in Table
\ref{tab:xi2potres}.  In the global fit to the off-axis potential at
$\beta=3.30$ we exclude the badly measured separations $2(2,1,0)$ and
$2(2,1,1)$.

\begin{table}[htbp]
\renewcommand{\arraystretch}{1.3}
  \begin{center}
    \begin{tabular*}{\textwidth}[c]{c@{\extracolsep{\fill}}cccccc}
      \hline\vspace{-0.05cm} 
      $\beta$ & gl. fit & $\alpha$ & $\sigma a_s a_t$ & loc. fit & $r_0/a_s$ & $r_0\sqrt{\sigma}$\\
      \hline
      3.00 & \phantom{ }$1-2\sqrt{6}$ & -0.201(18) & 0.3595(87) & $\sqrt{6}-3$ & 1.353(16) & 1.119(51)\\
      3.15 & $1-5$ & -0.1503(12) & 0.1654(6) & $1-3$ & 2.038(2) & 1.162(15)\\
      3.30 & $\sqrt{2}-5$ & -0.1539(7) & 0.0683(3) & $2\sqrt{2}-2\sqrt{3}$ & 3.154(9) & 1.140(8)\\
      3.50 & $1-6$ & -0.1478(4) & 0.0320(3) & $2-4$ & 4.906(21) & 1.189(14)\\
      \hline
    \end{tabular*}
    \caption{Results of the measurements of the static quark-antiquark potential using the $\xi=2$ perfect
      action. The parameters $\alpha$ and $\sigma$ of the potential ansatz
      eq.~(\ref{eq:isopot}) are given together with the
      global fit range chosen to determine them. The hadronic scale $r_0$ is given in spatial lattice units
      together with the local fit range used to determine that  
    quantity. Note, that we have chosen $c=6.00$ and $c=0.89$ for
      $\beta=3.0$ and $\beta=3.5$, respectively (see Section \ref{sec:xi2scale}). Additionally, the dimensionless quantity 
      $r_0\sqrt{\sigma}$ is given.}
    \label{tab:xi2potres}
  \end{center}
\end{table}

\begin{figure}[htbp]
\begin{center}
\includegraphics[width=10cm]{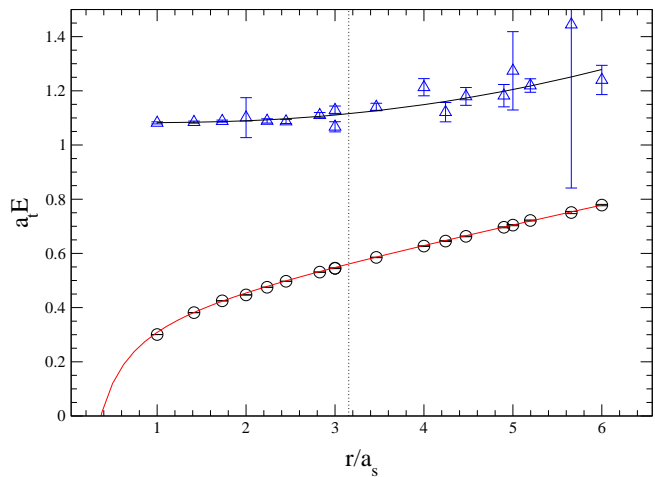}
\end{center}
\caption{Static quark antiquark potential at $\beta=3.3$ including the
first excited state $V^*(r)$ at $\beta=3.3$. The location of $r_0$ where the determination of the anisotropy is
done, using the gap, is marked by the dotted line.}
\label{pot_exst}
\end{figure}

In principle, it is possible to determine the renormalised anisotropy
using the first excited state of the potential together with the
ground state. The ground state values are again fitted to the ansatz
$V(r)=V_0+\alpha/r+\sigma r$, whereas the values of the excited state
are fitted to the (ad-hoc) ansatz $V^*(r)=A/r+B+Cr+Dr^2$. As can be
seen from Figure \ref{pot_exst}, however, for the off-axis potential
at $\beta=3.3$, the energy values of the first excited state have
large errors.  The result for the separation
$a_t(V^*(r_0)-V(r_0)) =0.555(47)$ thus shows a large error.  Comparing
this value to $r_0(V^*(r_0)-V(r_0))\approx 3.25(5)$ of
\cite{Morningstar:1998da} one obtains $\frac{r_0}{a_t}=5.86(59)$ and
thus $\xi_{\text{R}}=a_s/a_t$=1.86(19) for $\beta=3.3$ which is in
agreement with the value determined using the torelon dispersion
relation, $\xi_{\text{R}}$=1.912(9). A similar determination for the
on-axis potential at $\beta=3.50$ is even more difficult due to the
smaller number of separations measured. The result is $\xi_{\text{R}}$=1.66(32) which
does again agree with the torelon result,
$\xi_{\text{R}}=1.836(9)$. Due to the large errors (possibly caused by the
operators used, optimised for the ground state) this method of determining
the renormalised
anisotropy $\xi_{\text{R}}$ is not suitable in the context of this
work.  However, for the determination of the scale as well as of the
anisotropy on lattices with finer temporal lattice spacing (above all
with higher anisotropies) this way might be feasible at least if
off-axis separations are included in the measurement.

\subsubsection{Rotational invariance and scaling}
\begin{figure}[htbp]
\begin{center}
\includegraphics[width=10cm]{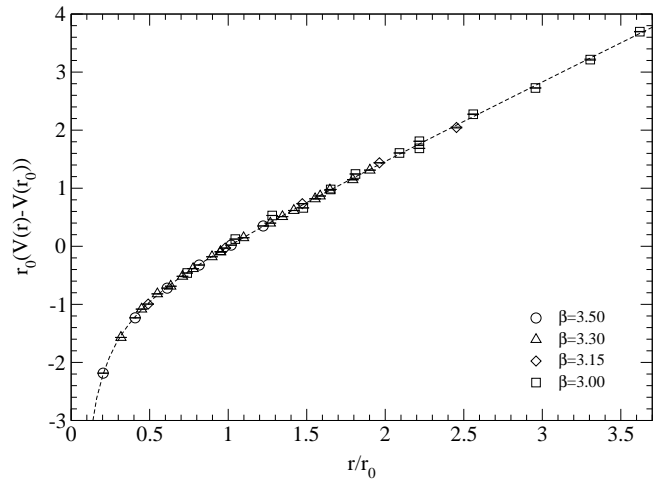}
\end{center}
\caption{The static quark-antiquark potential measurements expressed in RG invariant, dimensionless units $r/r_0$,
$r_0 V$. The unphysical constant $r_0 V(r_0)$ has been subtracted such that all the curves exactly coincide at $r=r_0$.
The dashed line is a global fit to the phenomenological potential ansatz, eq.~(\ref{eq:isopot}).}
\label{fig:allpot_xi2}
\end{figure}

The measurements at $\beta=3.0, 3.3$ including a large number of
off-axis separations provide information about the deviations from
rotational invariance. This issue has been addressed in
\cite{Blatter:1996ti} for the RGT
used in this work and  it turned out that violations of rotational
symmetry caused by the blocking are small. That this is still true for
the parametrisation employed here is shown by the off-axis potential at $\beta=3.30$,
see Figure \ref{fig:pot_grnd}. The deviations on the coarse lattice at
$\beta=3.00$ are significantly larger (see Figure
\ref{fig:allpot_xi2}) which is no surprise as this coupling
corresponds to a spatial lattice spacing $a_s\approx
0.37$~fm.
We present some dimensionless ratios containing
information about the deviation from rotational invariance in Tab. \ref{tab:rotinvpot}; due to the resulting
small values the errors are rather large. One notices that the
parametrised perfect action does a good job, for $\beta=3.30$ the deviation between the exactly degenerate
separations (2,2,1) and (3,0,0) is zero (within statistical errors).

\begin{table}[htbp]
\renewcommand{\arraystretch}{1.3}
  \begin{center}
    \begin{tabular*}{\textwidth}[c]{c@{\extracolsep{\fill}}cc}
      \hline\vspace{0.05cm}
      $\beta$ & $\frac{V(1,1,1)-V_{\text{fit}}(\sqrt{3})}{V(2,0,0)-V(1,0,0)}$ &
               $\frac{V(2,2,1)-V(3,0,0)}{V(3,0,0)-V(1,0,0)}$\\
      \hline\vspace{0.05cm}
      3.00 & 0.086(4) & 0.057(6)\\
      3.30 & 0.029(7) & 0.005(6)\\
      \hline
    \end{tabular*}
    \caption{The normalised deviations from rotational invariance for the AICP action.
 The ratios displayed are zero in the case of no rotational symmetry
        violation.}
    \label{tab:rotinvpot}
  \end{center}
\end{table}

The static $q\bar{q}$-potential is also an effective test of
scaling. Expressing the potential measurements performed at different
couplings $\beta$ in the RG invariant, dimensionless ratios $r/r_0$
and $r_0 V$ and subtracting the unphysical constant $r_0 V(r_0)$
should lead to potentials lying exactly on top of each
other. Deviations indicate either scaling violations or ambiguities in
the determination of $r_0$.  Figure \ref{fig:allpot_xi2} includes all
potential measurements and shows that the different curves can hardly
be distinguished from each other, except some energies determined on
the coarsest lattice at $\beta=3.00$. They deviate notably from the
curve which is a fit to the phenomenological potential ansatz, 
eq.~(\ref{eq:isopot}), including all the measurements that have been included
into the global fits of the single $\beta$ values, see Table
\ref{tab:xi2potres}. The results for the two physical parameters from
this global fit are $\alpha=-0.27799(5)$ and $\sigma r_0^2$=1.3690(3).

\subsection{The Critical Temperature}

The critical temperature of the deconfining phase transition contains
information about the temporal scale of the lattice at a given
coupling $\beta$ as $T_c=1/(N_t a_t)$. Comparing this information to
quantities obtained from the measurements of torelons or the static
quark-antiquark potential offers many interesting scaling
(and other) tests. However, since measurements of such quantities at the
determined values of $\beta_{\text{crit}}$ are yet absent and because
interpolations in $\beta$ are very difficult due to the
renormalisation of the anisotropy $\xi$, these tests are a future
project.

The critical temperature $T_c$ may be defined as the location of
the peak in the susceptibility of the Polyakov loop $P$ (Wilson line
in temporal direction),
\begin{equation}
\label{eq:susc}
\chi\equiv N_s^3 ( \langle |P|^2\rangle - \langle |P|\rangle^2 ).
\end{equation}

We determine the values of $\beta_c(N_t)$ on rather coarse lattices
with temporal extensions $N_t=3,\ldots,7$.
Simulations at different values of $\beta$ near the estimated critical
couplings $\beta_c$ are performed and the peaks are determined by
employing the Ferrenberg-Swendsen multi-histogram reweighting
\cite{Ferrenberg:1988yz,Ferrenberg:1989ui} (see also
\cite{Wenger:2000aa,Niedermayer:2000yx,Rufenacht:2001aa} for
additional details). The run parameters of the simulations are
collected in Table \ref{tab:xi2deconfrtp}.

In this first study, we decide not to examine the finite-size scaling
of $\beta_{\text{crit}}$ but to choose $L_s/L_t=\xi\cdot
N_s/N_t\approx 3.5\sim 4$. Obviously, the anisotropic nature of the
action makes the computational effort of the simulations for exploring
the deconfining phase transition smaller.  The effects of the finite
volume are (rather conservatively) estimated by using the finite-size
scaling relation for a first order phase transition,
\begin{equation}
\beta_c(N_t,N_s)=\beta_c(N_t,\infty)-h\left(\frac{N_t}{N_s}\right)^3.
\label{eq:fofss}
\end{equation}
The parameter $|h|$ is set to the largest value $|h|=0.25$ that has
been observed with the isotropic action, and the value of
$\xi_{\text{R}}$ used to estimate the spatial lattice volume is set to
the lowest value appearing in the whole range of $\beta$ values
considered.  The extrapolated critical couplings $\beta_c$ are given
together with the statistical error and the estimated finite-volume
error in the last column of Table \ref{tab:xi2deconfrtp}.

\begin{table}[htbp]
\renewcommand{\arraystretch}{1.3}
  \begin{center}
    \begin{tabular*}{\textwidth}[c]{c@{\extracolsep{\fill}}cccccc}
      \hline\vspace{-0.05cm}
      lattice size & $\beta$ & sweeps & $\tau_p$ & $\tau_{\text{int}}$ & $\chi_{\text{L}}$ & $\beta_{c}$ \\
      \hline
      $3\times 6^3$ & 2.80 & 45000 & & 70 & 1.45(5) & 2.863(1)(5)\\
                    & 2.85 & 45000 & 410 & 259 & 11.3(20) &\\
                    & 2.87 & 45000 & & 312 & 21.6(28) &\\
                    & 2.90 & 50000 & & 100 & 2.34(10) \\
      \hline
      $4\times 8^3$ & 3.00 & 45000 & & 81 & 1.63(5) & 3.032(1)(5)\\
                    & 3.02 & 45000 & 870 & 102 & 3.97(26) & \\
                    & 3.03 & 45000 & 830 & 372 & 43.2(22) &\\
                    & 3.04 & 45000 & & 188 & 10.9(18) &\\
                    & 3.05 & 55600 & & 144 & 3.94(21) &\\
      \hline
      $5\times 9^3$ & 3.07 & 44000 & & 117 & 0.84(4) & 3.118(1)(6)\\
                    & 3.10 & 45000 & 560 & 154 & 4.38(63) & \\
                    & 3.11 & 20500 & 640 & 198 & 15.0(27) & \\
                    & 3.12 & 44500 & 3700 & 278 & 33.2(19) & \\
                    & 3.15 & 20000 & & 110 & 3.26(18) & \\
      \hline
      $6\times 11^3$ & 3.17 & 45000 & 1440 & 121 & 7.4(11) & 3.181(1)(6)\\
                     & 3.18 & 45000 & 1750 & 201 & 30.0(15) &\\
                     & 3.185 & 45000 & & 232 & 23.2(24) & \\
                     & 3.19 & 45000 & & 123 & 10.4(14) &\\
                     & 3.20 & 45000 & & 100 & 5.62(36) &\\
      \hline
      $7\times 13^3$ & 3.22 & 32400 & 700 & 87 & 6.7(5) & 3.236(1)(6)\\
                     & 3.23 & 44500 & 750 & 129 & 18.4(11) & \\
                     & 3.24 & 29000 & & 149 & 20.7(16) & \\
                     & 3.25 & 22400 & & 83 & 9.2(10) \\
      \hline
    \end{tabular*}
    \caption{Run-time parameters of $\xi=2$ $T_c$-simulations. The number of sweeps, the persistence
time $\tau_p$ (if applicable), the integrated autocorrelation time $\tau_{\text{int}}$ and the Polyakov loop
susceptibility $\chi_{\text{L}}$ are given for all values of $\beta$ used in the final reweighting procedure,
as well as the resulting $\beta_c$. The systematic error in the second brackets is a (rather conservative)
estimate of the finite volume effects and has to be considered if the given value should be an estimate of
the infinite-volume critical coupling.} 
    \label{tab:xi2deconfrtp}
  \end{center}
\end{table}

\subsection{Glueballs}\label{sec:xi2gb}

\subsubsection{Introduction}

The particles mediating the strong interaction of QCD, the gluons,
carry colour charge and thus interact with each other, unlike, e.g.,
their counterpart in electromagnetism, the photons, which have zero
electric charge.  The spectrum of QCD may thus contain bound states of
(mainly) gluons, called glueballs.  These states are described by the
quantum numbers $J$ denoting the (integer) spin, $P$ denoting the
eigenvalue $\pm 1$ of the state under parity and $C$ denoting the
eigenvalue $\pm 1$ under charge conjugation. Thus the eigenstates of
the Hamiltonian corresponding to glueball states are labeled
$|J^{PC}\rangle$.

Currently, there is an ongoing debate whether light glueballs (above
all the scalar $0^{++}$ which is the lightest state in pure lattice
gauge theory with a mass of about 1.6~GeV) have been observed
experimentally at about the mass that is predicted by quenched
simulations on the lattice, whether the lightest glueball is much
lighter (below 1~GeV) and very broad \cite{Minkowski:1998mf}, or
whether glueballs have not been observed at all in
experiments \cite{Klempt:2000ud}. There are mainly two reasons for this
uncertainty.  On one hand, the experimental data seem not yet to be
accurate and complete enough, despite large efforts in the last years,
driven by the lattice results; on the other hand, lattice simulations
with high statistics, measuring glueball states, have been performed
only in the quenched approximation, where the quarks are infinitely
heavy and thus static. Decreasing the sea (dynamical) quark mass
(finally down to the physical value) will allow to track the glueball
states as sea quark effects are increased. It may turn out, that
indeed the glueball mass is lighter than the one measured in pure
gauge theory (for partially quenched results possibly indicating this
see \cite{Michael:1999rs,McNeile:2000xx,Bernard:2001av}). It may even
happen that by ``switching on'' the sea quarks the scalar glueball
starts to decay (almost) instantaneously
to $q\bar{q}$ states, i.e.~it ceases to exist physically.  However,
the (partially) unquenched results are rather indecisive yet.

\subsubsection{Glueballs on the Lattice}\label{sec:latgb}

Glueballs in the continuum are rotationally invariant and have a
certain (integer) spin $J$. On the lattice, the rotational $O(3)$
symmetry is broken, only its discrete cubic subgroup $O_h$ survives
the discretisation. Therefore, the eigenstates of the transfer matrix
are classified according to the five irreducible representations of
$O_h$: $A_1$, $A_2$, $E$, $T_1$, $T_2$ with dimensions 1, 1, 2, 3, 3
respectively. Their transformation properties may be described by
polynomials in the components $x$, $y$, $z$ of an $O(3)$ vector as
follows: $A_1\sim\{1\}$, $A_2\sim\{xyz\}$, $E\sim\{x^2-z^2,y^2-z^2\}$,
$T_1\sim\{x,y,z\}$, $T_2\sim\{xy,xz,yz\}$.  Generally, an $O(3)$
representation with spin $J$ splits into several representations of
the cubic group. Since $O_h$ is a subgroup of $O(3)$, any
representation $D_J$ with spin $J$ in the continuum induces a
so-called subduced representation $D_J \downarrow O_h$ on the
lattice. This subduced representation no longer has to be irreducible
but is a direct sum of irreducible representations $\Gamma^p$ of
$O_h$:
\begin{equation}
D_J \downarrow O_h = \Gamma^1 \oplus \Gamma^2 \oplus \cdots.
\end{equation}
Table \ref{tab:subdrep} lists the subduced representations of $D_J$
for $J=1,\ldots,6$. The spin $J=2$ state for example splits up into
the 2-dimensional representation $E$ and the 3-dimensional
representation $T_2$. Approaching the continuum, rotational symmetry
is expected to be restored and thus, as a consequence the mass
splitting of these two states will disappear and the two
representations form together the 5 states of a spin $J=2$ object.

\begin{table}
\begin{center}
\renewcommand{\arraystretch}{1.5}
\begin{tabular*}{\textwidth}{c@{\extracolsep{\fill}}ccccccc}
\hline
$\Gamma^p$ & $D_0$ & $D_1$ & $D_2$ & $D_3$ & $D_4$ & $D_5$ & $D_6$\\\hline
$A_1$ & 1 & 0 & 0 & 0 & 1 & 0 & 1\\
$A_2$ & 0 & 0 & 0 & 1 & 0 & 0 & 1\\
$E$   & 0 & 0 & 1 & 0 & 1 & 1 & 1\\
$T_1$ & 0 & 1 & 0 & 1 & 1 & 2 & 1\\
$T_2$ & 0 & 0 & 1 & 1 & 1 & 1 & 2\\
\hline
\end{tabular*}
\end{center}
\caption{The composition of the subduced representation $D_J \downarrow O_h$ in terms of the irreducible representations of the
cubic group $O_h$.}
\label{tab:subdrep}
\end{table}

Pure glue physical states on the lattice are created and annihilated
by applying gauge invariant operators to the pure gauge vacuum. In our
simulations, we use space-like Wilson loops in the fundamental
representation of SU(3). Since we do not aim at measuring non-zero
momentum glueballs we consider only translationally invariant
operators, i.e.~operators averaged in space.

It is computationally feasible to measure Wilson loop operators up to length
8. The composition of the irreducible representations $\Gamma^{PC}$ of the
cubic group in terms of these 22 loop shapes has been done already in Ref.
\cite{Berg:1983kp}.  Operators contribute to the two- and three-dimensional
representations with two and three different polarisations, respectively, in
analogy to different magnetic quantum numbers $m$ for a given angular momentum
$l$ in the O(3) group.  Measuring all these polarisations may suppress
statistical noise more than just increasing statistics since the different
polarisations of a loop shape are expected to be anti-correlated.

Due to mixing with other states present in pure gauge theory on a
periodic lattice, such as torelons or glueball pairs, and due to the
breaking of the continuum rotational symmetry by the lattice, the
identification and the continuum spin assignment of single glueball
states are additional vital questions which will be addressed in
Sections \ref{sec:gbident}, \ref{sec:spinident}. 

\subsubsection{Simulation Parameters}

To allow for a continuum extrapolation, at least for the lighter
states, we decide to perform simulations at three different lattice
spacings in the range 0.10~fm~$\leq a_s \leq$~0.25~fm in volumes
between (1.2~fm)$^3$ and (2.0~fm)$^3$. The simulation parameters are
given in Table \ref{tab:xi2gbpar}.  The gauge fields are updated by
performing compound sweeps consisting of 4 pseudo over-relaxation and
1 Metropolis sweep, and after every 2 compound sweeps we measure all
22 loop shapes up to length 8. The measurement is performed on
APE-smeared configurations (see Section \ref{sec:smearing}) in order to
spatially enlarge the operators, thus improving the overlap with the
glueball states and reducing unphysical high-momentum fluctuations.
On the coarse lattice at $\beta=3.15$ we use smearing levels ${\mathcal
S}_n, n = 2, 4, 6, 8, 10$, on the other lattices we use levels ${\mathcal
S}_n, n = 3, 6, 9, 12, 15$, always with smearing parameter
$\lambda_s=0.1$.

\begin{table}[htbp]
\renewcommand{\arraystretch}{1.3}
  \begin{center}
    \begin{tabular*}{\textwidth}[c]{c@{\extracolsep{\fill}}cccc}
      \hline\vspace{-0.05cm} 
      $\beta$ & $(a_s/r_0)^2$ & $S^3\times T$ & $V$ & \# sweeps / measurements\\
      \hline
      3.15 & 0.2408(5) & $8^3\times 16$ & (1.96~fm)$^3$ & 241000 / 24100\\
      3.30 & 0.1005(6) & $10^3\times 20$ & (1.59~fm)$^3$ & 99000 / 9900\\
      3.50 & 0.0415(4) & $12^3\times 24$ & (1.22~fm)$^3$ & 115000 / 11500\\
      \hline
    \end{tabular*}
    \caption{Run parameters for the glueball simulations using the $\xi=2$ perfect
      action.}
    \label{tab:xi2gbpar}
  \end{center}
\end{table}

\subsubsection{Determination of the Energies}

The masses of the lowest states and the first excited states (if
possible) in all the representations are determined using the
variational techniques described in Appendix B of
\cite{Niedermayer:2000yx}.  The results in units of the temporal
lattice spacing $a_t$ are listed in Tables \ref{tab:b315_fit_results}
-- \ref{tab:b350_fit_results} in Appendix \ref{app:rescoll} together
with the number $N$ of operators entering into the fitting procedure,
the time slices on which the initial generalised eigenvalue problem is
solved (usually 1/2), the number of operators $M$ in the final fit and
the corresponding value of $\chi^2/N_{\text{DF}}$. These results are
then multiplied by $\xi_R\cdot r_0/a_s$ from Tables~\ref{tab:xi2torres}
and \ref{tab:xi2potres} to obtain the glueball masses
in units of the hadronic scale $r_0$, listed in Table
\ref{tab:gb_r0_results}.

In order to obtain reliable estimates of the glueball masses from the
variational method, one has to pay attention that among the large
number of operators (up to 145 from five smearing levels) there are no
correlators entering the process that are measured exceptionally bad,
i.e.~with large errors even at small time separation, as this may
destabilise the determination. To filter out such data, we look at the
relative errors of the single operators depending on the temporal
separation of the creation and annihilation operators and drop all
correlators whose signal falls below a certain treshold value.  Usually,
this is repeated with different treshold values,
yielding different sized sets of operators ($2\sim 4$ sets per
representation).  We perform $O(100)$ mass determinations on each
operator set, varying the analysis parameters such as $M$ (see
Appendix B of \cite{Niedermayer:2000yx}) or the fitting range $t = t_{\text{min}} \ldots
t_{\text{max}}$.  It turns out that generally the variational
method
used is stable, provided that the badly measured operators are absent
from the beginning and provided that no important operators are
missing. Patently, to find such a window of the number of initial
operators, $N$,  may be difficult if there is a rather small total number of
operators.

For some of the heavy states, such as the $PC=--$ glueballs, whose masses can
only be determined from the $\beta=3.50$ measurements, we have to resort to
using the correlators at $t_0=0$ and $t_1=1$ for the solution of the first
generalised eigenvalue problem. This has the advantage that the correlators
from separation $t_0=0$ are positive definite by definition (which is not true
for $t_0\ge 1$), thus rendering the second generalised eigenvalue problem well
defined even if all the initial operators are kept, $M=N$. However, the
correlators $C(t=0)$ are under suspicion of containing rather little physical
information about the correlation lengths (which correspond to the
\emph{decay} of the correlators), actually they almost only provide
information about the relative normalisation of operators. Obviously,
operators with a large signal at $t=0$ might be preferable because their
signal/noise ratio is probably better on average, however in general
they do not correspond to the operators having the largest overlap with the
lowest-lying states. It turns out that one has to pay much more attention in
choosing the fit range $t=t_{\text{min}} \ldots t_{\text{max}}$ when using
$t_0=0$, $t_1=1$ because in this case at small $t$ the contamination of the
lowest masses due to higher states is much more significant.

During the analysis, it turns out that the mass determinations for the
scalar representation $A_1^{++}$ are rather challenging. The ground
state receives a larger relative error than comparable states where
the operators are measured equally well; for the first excited state
it is very hard to obtain a stable determination, and the error of its
mass may turn out to be huge (e.g. at $\beta=3.50$). Probably, it is
the underlying vacuum, having the same quantum numbers as the glueball
state, that is responsible for these troubles. We treat the vacuum
just like another state in this representation, so that the glueball
ground state is effectively a first excited state and the glueball
first excited state is effectively a second excited state of the
representation. Attempts of using other ways of getting rid of the
vacuum such as the usual v.e.v.~subtraction or the subtraction of large $t$
correlators (that are assumed to contain solely noise, however
correlated to the noise at lower $t$) or even more sophisticated
methods (like solving a generalised eigenvalue problem using large $t$
correlators to dig out the vacuum state) do not at all succeed in
improving the situation.

\begin{table}[htbp]
  \begin{center}
    \renewcommand{\arraystretch}{1.5}
    \begin{tabular*}{\textwidth}{c@{\extracolsep{\fill}}cccc}
      \hline
      Channel & $J$ & $\beta=3.15$ & $\beta=3.30$ & $\beta=3.50$\\
      \hline
      $A_1^{++}$   & 0 & 2.58(9) & 3.55(13) & 3.65(15)\\
    ${A_1^{++}}^*$ & 0 & 5.47(53) & 6.83(37) & 6.49(148)\\
      $E^{++}$     & 2 & 5.63(23) & 5.93(15) & 6.08(28)\\
      ${E^{++}}^*$ & 2 & 7.95(73) & 8.76(48) & 10.66(59)\\
      $T_2^{++}$   & 2 & 5.67(25) & 5.82(13) & 6.13(16)\\
    ${T_2^{++}}^*$ & 2 & & 7.97(69) & 10.16(30)\\
      $A_2^{++}$   & 3 & 6.71(73) & 9.11(64) & 10.92(48)\\
    ${A_2^{++}}^*$ &   & & & 14.12(130)\\
      $T_1^{++}$   & 3 & 7.54(68) & 9.00(44) & 11.21(33)\\
    ${T_1^{++}}^*$ &   & 8.69(91) & &\\
\hline
      $A_1^{-+}$   & 0 & 6.36(59) & 6.13(28) & 6.79(27)\\
    ${A_1^{-+}}^*$ &   & & & 10.27(110)\\
      $E^{-+}$     & 2 & 7.41(48) & 8.23(28) & 8.48(29)\\
      ${E^{-+}}^*$ &   & 8.35(125) & & 12.66(84)\\
      $T_2^{-+}$   & 2 & 7.41(49) & 8.18(21) & 8.57(42)\\
    ${T_2^{-+}}^*$ &   & & & 15.08(67)\\
\hline
      $T_1^{+-}$   & 1 & 7.20(39) & 7.69(26) & 9.12(24)\\
      $A_2^{+-}$   & 3 & & & 11.02(68)\\
    ${A_2^{+-}}^*$ &   & & & 15.55(147)\\
      $T_2^{+-}$   &   & 7.75(87) & & 10.29(118)\\
    ${T_2^{+-}}^*$ &   & & & 14.98(139)\\
     $E^{+-}$      &   & & & 12.42(42)\\
\hline
     $T_1^{--}$    &   & & & 10.81(83)\\
     $T_2^{--}$    &   & & & 10.85(83)\\
     $A_2^{--}$    &   & & & 12.18(45)\\
     $A_1^{--}$    &   & & & 15.34(103)\\
    \hline
    \end{tabular*}
    \caption{{}Final glueball mass estimates in terms of the hadronic scale $r_0$, $m_G r_0$ from the measurements
using the perfect $\xi=2$ action. The continuum spin assignment $J$ is given as well.}
    \label{tab:gb_r0_results}
  \end{center}
\end{table}

\subsubsection{Torelons and Multi-Glueball States}\label{sec:gbident}

Besides single glueballs, the spectrum of pure gauge theory on a
lattice with periodic boundary conditions also contains states
consisting of several glueballs, torelons or mixed states of glueballs
and torelons. Although we expect that the operators used to measure
glueball energies couple most strongly to the single glueball states,
other states with similar masses and compatible quantum numbers might
mix with them and distort the result, i.e.~the energy determined by
the analysis of the correlation matrix may be dominated by a
multi-glueball or torelon state with smaller energy than the
single-glueball state to be measured.

In principle, there are several means of determining the nature of a
state that has been measured.
Firstly, the simulation may be repeated
on lattices of different physical size, keeping the lattice spacing fixed. As
multi-glueball and torelon states show a finite-volume scaling behaviour very
different from single glueballs, such states stand out and may be
dropped from further analysis. 
Secondly, one may measure additional
operators that couple strongly to torelon or multi-glueball
states. Including or excluding these operators in the variational
method and studying the coefficients obtained from the variational method,
the mixing strength is determined and states which do not mix
considerably with any of the additional operators may be safely
considered to be single-glueball states.

Finally, using the mass
estimates for the low-lying glueballs, one may determine the
approximate locations of the lowest-lying multi-glueball states. Also,
the minimal energy of mixing torelons may be estimated using the
string formula, eq.~(\ref{eq:stringtorelon}).

The first two of the methods mentioned above require additional work
and computer time and go beyond the scope of this work. The last
method, however, can be done rather easily. Let us first calculate the
energy of the lowest-lying torelon states that may interfere with our
measurements. Single torelons (see Section \ref{sec:xir}) transform
non-trivially under $Z_3$ symmetry operations; our operators, closed
Wilson loops, however, are invariant under these transformations. This
means that they cannot create single torelon states, however the
creation of two torelons of opposite center charge is possible. If we
assume that they do not interact considerably and if our lattice
extension is rather large then we can use the simple formula
$E_{2T}\approx 2\sigma L$ to estimate the energy of a torelon pair
with momentum zero, where $\sigma$ is the string tension (measured by
the static $q\bar{q}$ potential, see Section \ref{sec:xi2scale}) and
$L$ is the spatial extent of the lattice. Table \ref{tab:min2tln}
lists the minimum energies of torelon pairs to be expected on our
lattices.

\begin{table}[htbp]
\begin{center}
  \renewcommand{\arraystretch}{1.5}
  \begin{tabular*}{4cm}{c@{\extracolsep{\fill}}c}
    \hline
    $\beta$ & $r_0 E_{2T}$\\
    \hline
    3.15 & 10.60\\
    3.30 & 8.24\\
    3.50 & 6.92\\
    \hline
  \end{tabular*}
  \caption{{}The minimum energies of momentum zero torelon pairs on
the lattices used in the glueball simulations, given in units of
$r_0^{-1}$.}
  \label{tab:min2tln}
\end{center}
\end{table}

Note, that a state composed of two opposite center charge torelons and
with total zero momentum is symmetric under charge conjugation. The
operators for $C=-$ states therefore do not create such torelon pairs.

To estimate the lowest energies of multi-glueball states present on
our lattices, we follow the method used by Morningstar and Peardon
\cite{Morningstar:1999rf}, described in great detail in Appendix F of
\cite{Rufenacht:2001aa}.  We assume that the energy of multi-glueball
states is approximately given by the sum of the energies of the
individual glueballs, i.e.~that there is no substantial
interaction. It is therefore clear that the lowest-lying
multi-glueball states with zero total momentum are the two-glueball
states, with energy
\begin{equation}
E_{2G}\approx\sqrt{\vec{p}^2+m_1^2}+\sqrt{\vec{p}^2+m_2^2},
\end{equation}
where $m_1$ denotes the rest mass of the first glueball with momentum
$\vec{p}$ and $m_2$ denotes the rest mass of the second glueball with
momentum $-\vec{p}$. The masses are taken from our determinations at
the respective values of $\beta$ and the (lattice) momenta are chosen
such that the energy $E_{2G}$ of the glueball pair contributing to
some representation $\Gamma^{PC}$ is minimised. Note, that glueball
pairs contribute to different representations of the cubic group, not
only depending on the representations according to which the single
glueballs transform, but also on the single glueball momentum
$\vec{p}$. 

\begin{figure}[htbp]
\begin{center}
\includegraphics[width=10.5cm]{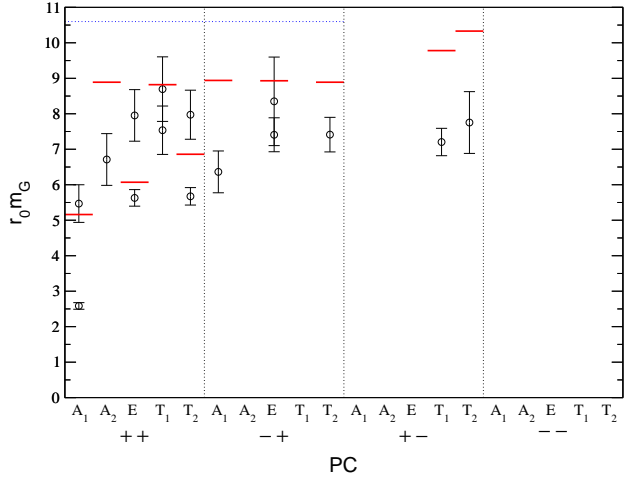}
\end{center}
\caption{The results of the glueball measurements at $\beta=3.15$ converted into units of $r_0$. The lower bounds
of the energies of multi-glueball states (solid lines) or torelon pairs (dotted line) are given as well.}
\label{fig:gbres_b315}
\end{figure}

\begin{figure}[htbp]
\begin{center}
\includegraphics[width=10.5cm]{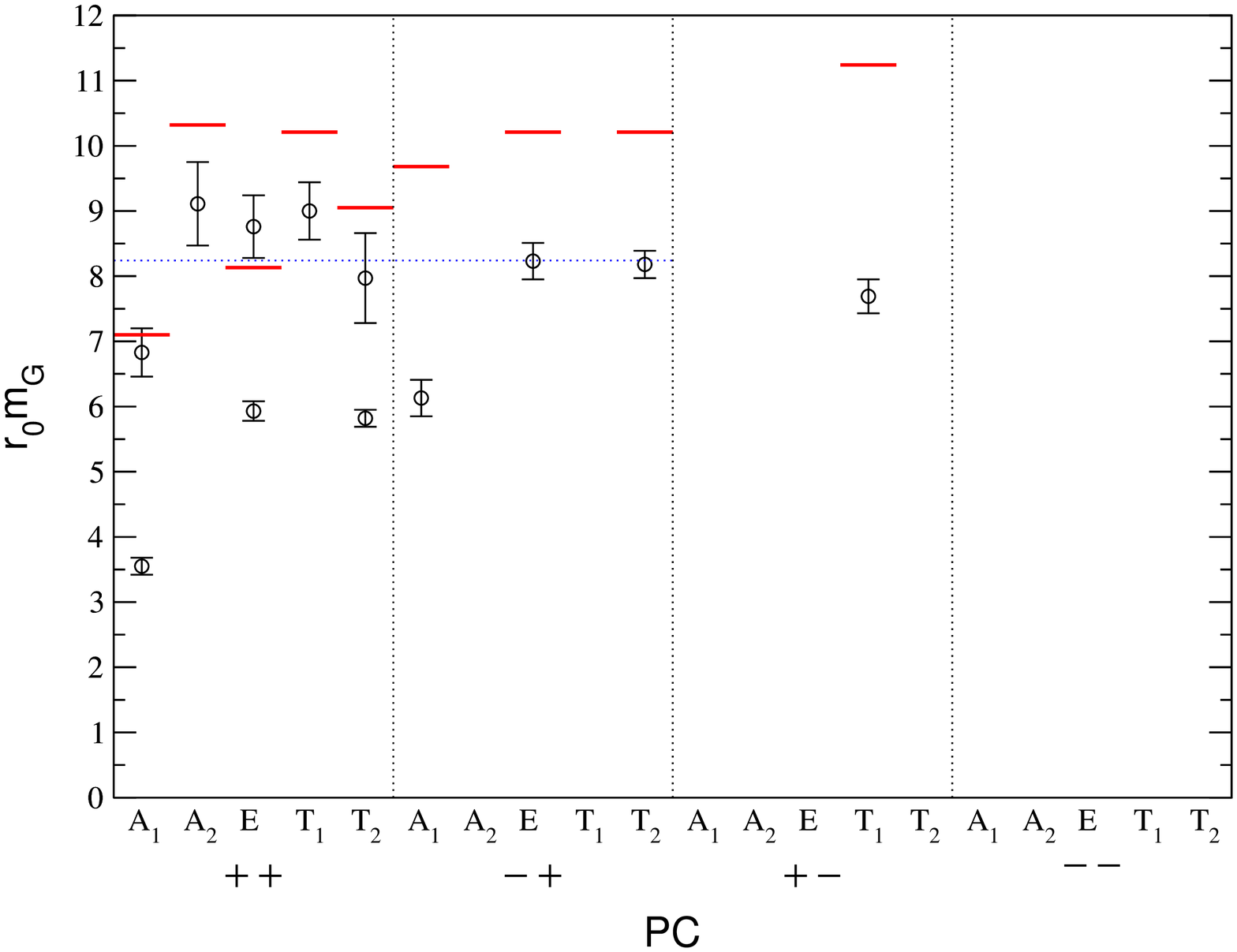}
\end{center}
\caption{The results of the glueball measurements at $\beta=3.30$ converted into units of $r_0$. The lower bounds
of the energies of multi-glueball states (solid lines) or torelon pairs (dotted line) are given as well.}
\label{fig:gbres_b330}
\end{figure}

\begin{figure}[htbp]
\begin{center}
\includegraphics[width=10.5cm]{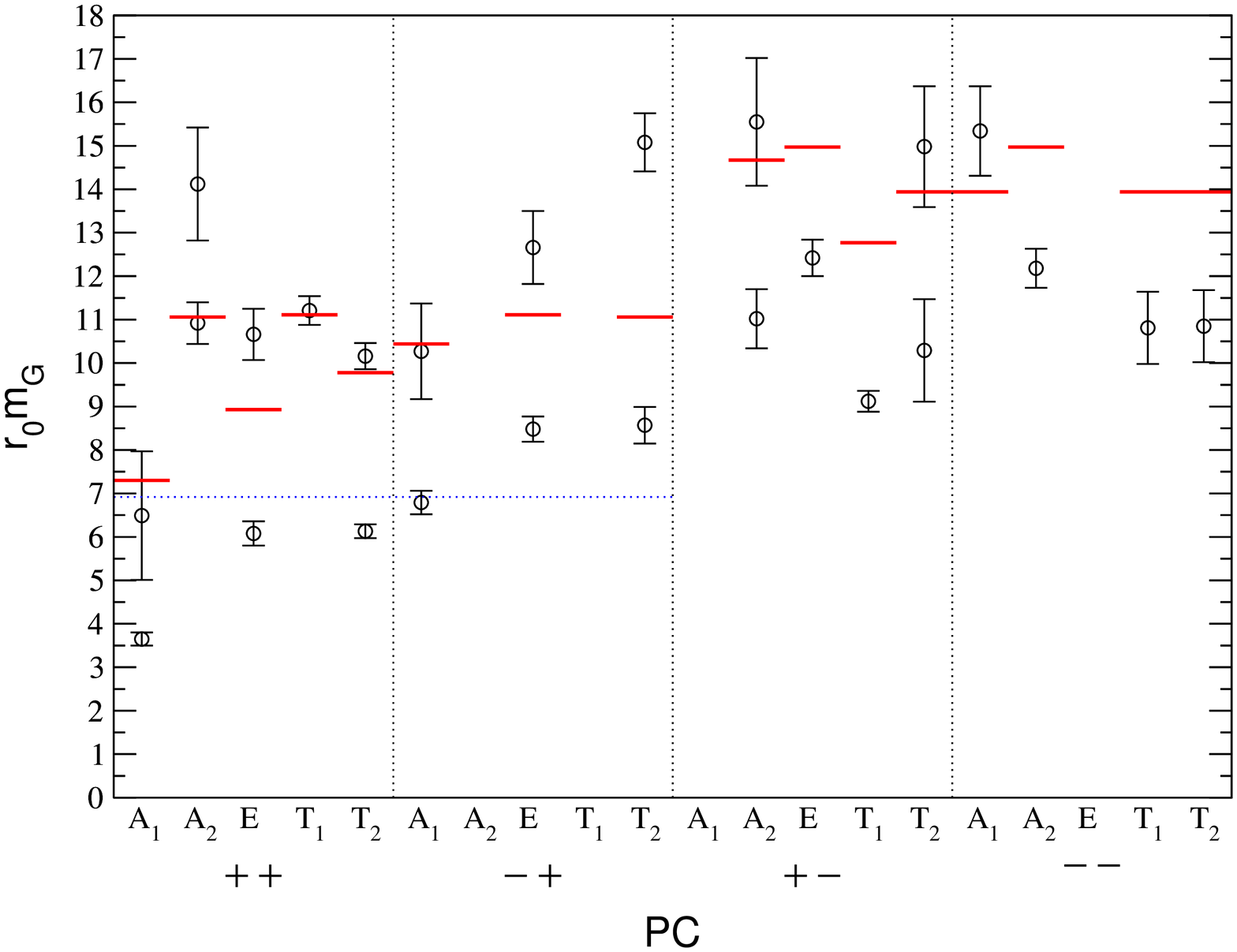}
\end{center}
\caption{The results of the glueball measurements at $\beta=3.50$ converted into units of $r_0$. The lower bounds
of the energies of multi-glueball states (solid lines) or torelon pairs (dotted line) are given as well.}
\label{fig:gbres_b350}
\end{figure}

The lower bounds of the multi-glueball energy region as well as the
lower bound for torelon pairs are indicated in Figures
\ref{fig:gbres_b315}--\ref{fig:gbres_b350}, together with the
determined energies of all the states in the respective
representations. It turns out that all of the excited states measured
and even some of the lowest states in a given representation could be
affected by torelon pairs or multi-glueball states. Of course, it is
not at all ruled out that these states are indeed single glueballs,
however one has to be very careful with the interpretation and has to
keep in mind that these issues require further study.

\subsubsection{The Continuum Limit}

In order to remove discretisation errors from glueball masses $r_0 m_G$
obtained at finite lattice spacing, we have to perform the continuum limit,
i.e.~to extrapolate the results to $a_s=0$. Having obtained results at three
values of the coupling $\beta$, one of them corresponding to a rather coarse
lattice spacing, and having no accurate information (e.g.~from perturbation
theory) about the behaviour of the energies depending on the cut-off
(about the form of the curve used in the extrapolation) this is not an easy
task and rather ambiguous.

We perform two different kinds of continuum limits. Firstly, we
extrapolate the masses in terms of the hadronic scale $r_0$, $r_0 m_G
= m_G a_t \cdot \xi \cdot r_0/a_s$, secondly, we extrapolate mass
ratios $m_{G_1}/m_{G_2}$ of different glueball representations.
Generally, the procedure we resort to is the following: We include
all the three energies or mass ratios, obtained at $\beta=3.15, 3.30,
3.50$, into the fit and use the form
\begin{equation}\label{eq:gbfitform2}
r_0 m_G\mid_{a_s} = r_0 m_G\mid_{a_s=0} + c_2 \left(\frac{a_s}{r_0}\right)^2,
\end{equation}\
thus including the correction of the smallest power of the (spatial) lattice
spacing $a_s$ to be expected to occur.
Additionally, we perform fits including only the results from the two
finer lattices at $\beta= 3.30, 3.50$ using a constant (if this fits
the results reasonably well) or the form stated above. This value of
the continuum mass (or mass ratio) is then compared to the result of
the fit of all the three data points, which is only accepted if the
two results coincide within their errors.

\subsubsection*{Masses in Terms of $\mathbf{r_0}$}

For the extrapolation of masses in terms of the hadronic scale $r_0$,
this procedure works very well for the representations $A_2^{++}$,
$E^{-+}$, $E^{++}$, ${E^{++}}^*$, $T_2^{-+}$ and $T_2^{++}$ where we
may always fit the results from the two finer lattices to a constant
which agrees with the result of the extrapolation using
eq.~(\ref{eq:gbfitform2}) on all three data points. For the $A_1^{-+}$
representation, we obtain the best results from fits with a
constant to two as well as to three data points.

Much more difficult is the extrapolation for the scalar glueball and
its excited state, $A_1^{++}$ and ${A_1^{++}}^*$. It turns out that
the fits have bad values of $\chi^2$ and for the ground state determinations
the result obtained performing a fit to a constant mass using the results from the
two finer lattices does not correspond to the fit including the third,
coarser lattice and a quadratic term in $a_s$. We decide to keep the (usual) result,
including the quadratic term and keep in mind that there may be present some
ambiguities which will be discussed later.

\begin{figure}[htbp]
\begin{center}
\includegraphics[width=12cm]{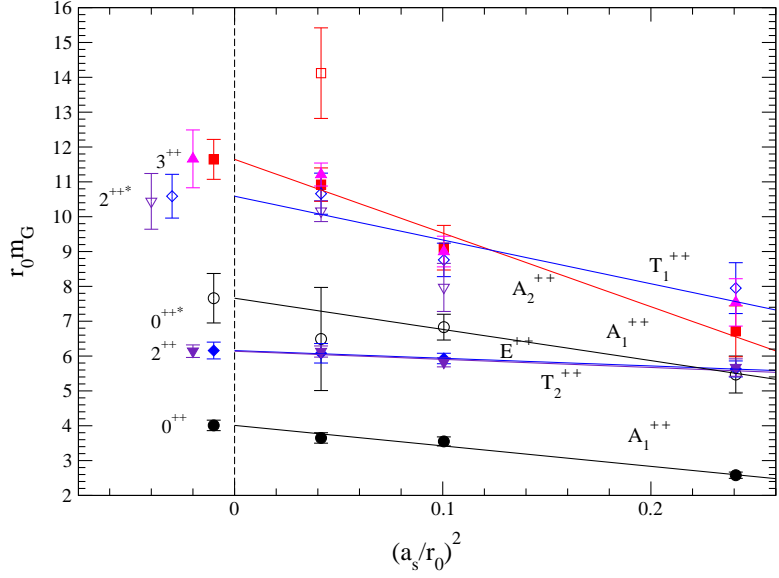}
\end{center}
\caption{Mass estimates of the $PC=++$ glueballs in terms of the
hadronic scale $r_0$ against $(a_s/r_0)^2$. The curves are the
continuum limit extrapolations of the forms as indicated in the text. \emph{Circles}: $A_1^{++}$, \emph{boxes}:
$A_2^{++}$, \emph{diamonds}: $E^{++}$, \emph{upward triangles}:
$T_1^{++}$, \emph{downward triangles}: $T_2^{++}$; \emph{solid
symbols}: ground states, \emph{open symbols}: first excited
states. Note that the continuum results of the representations
$T_1^{++}$ and ${T_2^{++}}^*$ have been obtained using a fit of
glueball mass ratios.}
\label{fig:gb_ex_pp}
\end{figure}

\begin{figure}[htbp]
\begin{center}
\includegraphics[width=12cm]{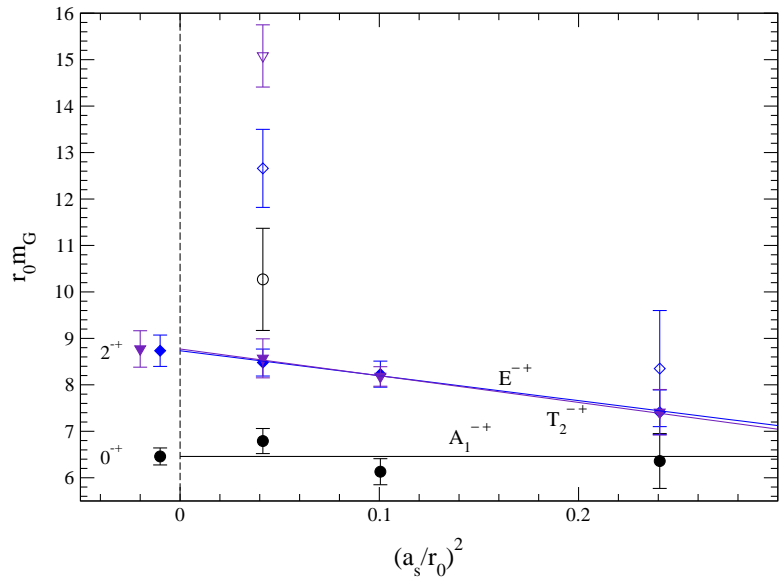}
\end{center}
\caption{Mass estimates of the $PC=-+$ glueballs in terms of the
hadronic scale $r_0$ against $(a_s/r_0)^2$. The curves are the
continuum limit extrapolations of the form $r_0
m_G+c_2(a_s/r_0)^2$. \emph{Circles}: $A_1^{-+}$, \emph{diamonds}:
$E^{-+}$, \emph{downward triangles}: $T_2^{-+}$; \emph{solid symbols}:
ground states, \emph{open symbols}: first excited states.}
\label{fig:gb_ex_mp}
\end{figure}

The results
of the continuum extrapolations for various glueball states are collected in
Table \ref{tab:cont_ex}. In Figures \ref{fig:gb_ex_pp} and \ref{fig:gb_ex_mp}
the fitting curves are shown together with the measured values for the $PC=++$
and the $PC=-+$ sectors where extrapolations are possible with our
data.

Other representations where energies can be determined at several
values of $\beta$, such as $T_1^{+-}$, $T_1^{++}$ or ${T_2^{++}}^*$, do
not yield consistent results using the procedure presented above and
we decide to abstain from performing continuum extrapolations in terms
of $r_0 m_G$ for these states.

\subsubsection*{Mass Ratios}

Extrapolating glueball mass ratios $m_{G_1}/m_{G_2}$ to the continuum has the
advantage that uncertainties present in the determination of $r_0/a_s$ and the
renormalised anisotropy $\xi_R$ cancel out.  Furthermore, we may assume that
finite size effects are similar for glueball states measured on the same
lattice as opposed to completely different quantities used e.g.~to obtain $r_0/a_s$
and $\xi_R$.

The common way of extrapolating glueball mass ratios to the continuum
is using ratios $m_G/m_{0^{++}}$, i.e.~ratios to the scalar glueball
mass, which is the lightest mass present. Doing this, we observe that
the extrapolation is difficult, the value of $\chi^2$ of the fit is rather bad
and the errors are large.  Furthermore, the resulting continuum mass
ratios are not in good agreement to ratios of the continuum masses
extrapolated in terms of $r_0$. This is
not a surprise, because the behaviour of the scalar glueball as a
function of the lattice spacing is rather complicated, i.e.~there are
large cut-off effects that cannot be described too
easily. Additionally, the errors of the mass estimates of the scalar
glueball are quite large. Due to these reasons, we decide to use the
well measured mass of the tensor $T_2^{++}$ representation, which
show small errors and seem to scale rather well.

The procedure of the extrapolation is the same as in the case of $r_0
m_G$ described in the previous section; the unambiguous results are
listed in Table \ref{tab:cont_ex_ratio}.  Again, for the $A_1^{-+}$
states, as well as for the $E^{++}$, the fit to a constant works out
best. Note,
that in this way we can obtain reliable continuum results for the representations
$T_1^{+-}$, $T_1^{++}$ and ${T_2^{++}}^*$ which is not possible
performing the extrapolation using masses in terms of
$r_0$. Furthermore, all the other results listed are in agreement to
the ratios of the continuum masses $r_0 m_G$ within one standard
deviation.

In the following, we will use the finite lattice spacing results from the
simulation on the finest lattice at $\beta=3.50$ for the energies of states
that cannot be extrapolated to the continuum in one of the two ways, e.g.~to
study degeneracies or to draw a (rather qualitative) picture of the low-lying
glueball spectrum. In tables listing continuum results, the values not
extrapolated to $a_s=0$ will be stated in square brackets.

\subsubsection{Continuum Spin Identification}\label{sec:spinident}
Once the extrapolation of the measured glueball  energies to the
continuum, $a_s=0$, has been performed, what remains is the assignment
of continuum spin $J$ to the different representations of the cubic
group $\Gamma$.  From Table \ref{tab:subdrep} we know from which
lattice representations $\Gamma^{PC}$ a continuum glueball with
quantum numbers $J^{PC}$ may obtain contributions, where the
assignment of parity $P$ and charge conjugation $C$ is simply
one to one. Degeneracies (in the continuum limit) between several
representations $\Gamma^{PC}$ contributing to the same continuum state
$J^{PC}$ are a strong indication for the correctness of the assignment
of all the lattice states involved to the same continuum $J^{PC}$
glueball. Furthermore, we make the assumption that the mass of the
glueballs increases with the spin getting larger.

In the $PC=++$ sector, we observe a single low-lying state,
$A_1^{++}$, which is thus assigned $J=0$.  The $E^{++}$ and $T_2^{++}$
states are degenerate to a very high precision as it should happen if
these representations correspond to the five polarisations of a $J=2$
glueball. The excited state ${A_1^{++}}^*$ again has no degenerate
partner and is assigned to an excited state of the continuum $J=0$
glueball. The excited states ${E^{++}}^*$ and ${T_2^{++}}^*$ again
turn out to be degenerate and so they are assigned to an excitation of
the $J=2$ glueball. The same situation is met with the representations
$A_2^{++}$ and $T_1^{++}$ which are thus assigned $J=3$. Since the first
excitation ${T_2^{++}}^*$ is part of the $J=2$ state, the missing 3
polarisations should come from the second excitation of $T_2^{++}$
which, however, cannot be measured by our simulations.

In the $PC=-+$ sector, the situation of the $A_1^{-+}$ and the
$E^{-+}$ and $T_2^{-+}$ is very similar to the one of their partners
in the $PC=++$ sector; these states are thus assigned $J=0$ and $J=2$
respectively.  The remaining three, excited states have no apparent
degeneracies, there is thus no safe assignment of continuum spin to
these states. The facts that the excited state ${A_1^{-+}}^*$ is
rather light and comes with no degenerate partner, raises the
presumption that it is the excited $J=0^{-+}$ glueball.

In the other two sectors, continuum extrapolations of our measured
energies are not possible, except for the representation $T_1^{+-}$
using mass ratios. However, looking at the degenerate representations
in the $PC=++, -+$ sectors, it is noticeable that the degeneracies are
apparent even for finite lattice spacing at $\beta=3.50,
3.30$. Assuming that this behaviour persists for $PC= +-, --$, at
least for energies not too high, we may assign continuum spin even to
some of the remaining states.

In the $PC=+-$ sector, we notice the lowest-lying $T_1^{+-}$ state
having no degenerate partner thence suggesting a $J=1$
interpretation. Next, there is the $A_2^{+-}$ which is likely to
correspond to $J=3$, which is the smallest possible spin corresponding
to this representation.

Finally, in the $PC=--$ sector, there are different possible
scenarios. The almost exact degeneracy of $T_1^{--}$ and $T_2^{--}$
suggests them contributing to the $J=3$ continuum state. The rather
heavier state $A_2^{--}$ could still be degenerate with the two latter
states (note the high mass and the finite lattice spacing!) and carry
the remaining polarisation. Admittedly, it could as well correspond to
an excited $J=3$ state or even to $J=6$.  The final $A_1^{--}$ state
is very heavy and indicates that glueballs with even spin and $PC=--$
have large energies as the representation $A_1$ contributes solely to
even numbered spin states up to $J=8$. 

Let us remark that a new method of improving the overlap of operators to
glueballs of a given continuum spin $J$ thus rectifying the continuum spin
identification is presented in \cite{Liu:2001je,Liu:2001wq}.

The well supported continuum spin assignments presented above
are given together with the glueball masses in Tables
\ref{tab:gb_r0_results}, \ref{tab:cont_ex}, \ref{tab:cont_ex_ratio}, while
the final results on the masses of these states are collected,
together with the masses in MeV, in Table \ref{tab:final_gb}.

\begin{table}[htbp]
  \begin{center}
      \renewcommand{\arraystretch}{1.5}
      \begin{tabular*}{\textwidth}{l@{\extracolsep{\fill}}lcl} \hline
      $J^{PC}$ & $\Gamma^{PC}$ & $m_G r_0$ & $m_G$~[MeV]\\ \hline
      $0^{++}$ & $A_1^{++}$ & 4.01(15) & 1580(60)\\ ${0^{++}}^*$ &
      ${A_1^{++}}^*$ & 7.66(71) & 3030(280)\\ $2^{++}$ & $E^{++}$,
      $T_2^{++}$ & 6.15(15) & 2430(60)\\ ${2^{++}}^*$ & ${E^{++}}^*$,
      ${T_2^{++}}^*$ & 10.52(51) & 4160(200)\\ ${3^{++}}$ &
      $A_2^{++}$, $T_1^{++}$ & 11.65(49) & 4600(190)\\ ${0^{-+}}$ &
      $A_1^{-+}$ & 6.46(18) & 2550(70)\\ ${2^{-+}}$ & $E^{-+}$,
      $T_2^{-+}$ & 8.75(26) & 3460(110)\\ ${0^{-+}}^*$ &
      ${A_1^{-+}}^*$ & [10.27(110)] & [4060(430)]\\ ${1^{+-}}$ &
      $T_1^{+-}$ & 9.45(71) & 3730(280)\\ ${3^{+-}}$ & $A_2^{+-}$ &
      [11.02(68)] & [4350(270)]\\ \hline \end{tabular*}
 \caption{{}Final results for the masses of continuum
      glueballs with quantum numbers $J^{PC}$, obtained from the
      lattice representations $\Gamma^{PC}$. For the conversion to
      MeV, $r_0^{-1}\approx 0.5$~fm~$\approx 395$~MeV has been
      used. Quantities in brackets have not been extrapolated to the
      continuum but denote results from the simulation on the finest
      lattice at $\beta=3.50$ (corresponding to $a_s=0.102$~fm).}
      \label{tab:final_gb} \end{center}
\end{table}

\subsubsection{Discussion}

In order to discuss the results obtained from the glueball
measurements let us first list
 the sources of problems that may affect our results. Firstly, there are finite
size effects; our finest lattice ($\beta=3.50$) has got a spatial
volume of (1.22~fm)$^3$ which is rather small compared to other
volumes used in glueball measurements. The magnitude of finite-size
effects depends largely on the quantity studied, more precisely on the
form of the glueball wave functions (above all their extension). Due to
the three volumes employed having such different size, finite-size
effects may not only systematically shift the lattice results but they
can also make reliable continuum extrapolations much more difficult or
even impossible.

Secondly, the mass of the scalar glueball $A_1^{++}$ (above all the
ground state) shows very large cut-off effects.  These could be due to
the presence of a critical end point of a line of phase transitions in
the fundamental-adjoint coupling plane. Our parametrisation includes
in its rich structure operators transforming in the adjoint
representation. If the net coupling of the parametrised action (which
we do not control during the construction and the parametrisation)
lies in a certain region, the effect of the critical end-point on
scalar quantities at certain lattice spacings (sometimes called the
``scalar dip'') may even be enhanced compared to other (more standard)
discretisations with purely fundamental operators. It is important to
note that the true classically perfect action is not expected to be
sensitive to this critical point, but it is rather the parametrisation
which may incidentally cause the sensitivity close to it.

Furthermore, our analysis shows that states consisting of several
glueballs or torelon pairs could mix with most of the higher-lying
states occurring in our measurements. This makes the interpretation of
the measured states more difficult, additionally, systematic effects
(e.g.~different mass shifts at different values of $\beta$ because of
the different lattice sizes or due to the presence of operators having
larger overlap with unwanted states) may again complicate the
continuum extrapolation.

There are several possibilities of improving the situation. Firstly, and
most simple, improving the statistics may help. This can be seen e.g.~by
looking at the results of the simulation on the coarsest lattice at
$\beta=3.15$.  Although the lattice is coarser, more energies can be
determined than on the finer $\beta=3.3$-lattice, mainly due to the larger
statistics. Additionally, a lot of problems stated previously could be kept
under control if for each value of $\beta$, we determine the glueball
energies on several lattices of different physical size.  A way of improving
the measurements is systematically studying the operators employed to create
and annihilate single glueball states in order to increase their overlap with
the single glueballs to be measured and to decrease their overlap with all the
other states. This includes the study of other smearing techniques, such as
e.g.~Teper fuzzing \cite{Teper:1987wt}. Conversely, the introduction of
operators coupling most strongly to the unwanted multi-glueball or
torelon-pair states allows for the calculation of mixing strengths in order to
exclude all the unwanted states from further analysis. Another technically
rather easy but computationally expensive way of improving the results is
performing simulations on additional values of $\beta$. This allows for a more
detailed study of the continuum limit, especially in the case of
non-perturbatively improved actions, where there is no clearcut information
about the way how the continuum is approached. Naturally, one has to take care
not to go too deep into the strong coupling region as there is not much
information about the continuum.

There is one error which is not cured by the measures recommended in
the previous paragraph, namely the effects on the scalar states, coming
from the ``scalar dip''. Morningstar and Peardon improve the situation
by adding a two-plaquette adjoint term with a negative coefficient which
results in an approach to the continuum on a trajectory always far
away from the dangerous ``dip'' region \cite{Morningstar:1999dh}. In
principle, the classically perfect action could be treated the same
way: extract all the operators with adjoint contributions present in
the parametrisation, determine their sign and add another (non-linear)
constraint to the fit, namely that the action of all these operators
together corresponds to an adjoint operator with a certain (negative)
coupling. Because of the large freedom in the fit, the inclusion of
this single criterion should not impair the quality of the
parametrisation considerably.

\begin{table}[htbp]
  \begin{center} \renewcommand{\arraystretch}{1.3}
      \begin{tabular*}{9cm}[c]{l@{\extracolsep{\fill}}ll}
      \hline\vspace{-0.05cm} Collab. & $r_0 m_{0^{++}}$ & $r_0
      m_{2^{++}}$\\ \hline UKQCD \cite{Bali:1993fb} & 4.05(16) &
      5.84(18) \\ Teper \cite{Teper:1998kw} & 4.35(11) & 6.18(21) \\
      GF11 \cite{Vaccarino:1999ku} & 4.33(10) & 6.04(18) \\ M\&P
      \cite{Morningstar:1999rf} & 4.21(15) & 5.85(8) \\ Liu
      \cite{Liu:2000ce} & 4.23(22) & 5.85(23) \\ \hline FP action &
      4.12(21) & [5.96(24)] \\ AICP & 4.01(15) & 6.15(15) \\ \hline
      \end{tabular*}
     \caption{{}Comparison of the lowest-lying glueball masses in units
      of $r_0$.  Values in brackets denote masses obtained at a
      lattice spacing $a=0.10$ fm and are not extrapolated to the
      continuum.}
\label{tab:cmpgbcont} \end{center}
\end{table}

\begin{table}[htbp]
  \begin{center}
    \renewcommand{\arraystretch}{1.3}
    \begin{tabular*}{\textwidth}[c]{l@{\extracolsep{\fill}}llll}
      \hline\vspace{-0.05cm}
    Collab. & $r_0 m_{0^{-+}}$ & $r_0 m_{{0^{++}}^*}$ & $r_0 m_{2^{-+}}$ & $r_0 m_{1^{+-}}$\\
      \hline
      Teper \cite{Teper:1998kw}       & 5.94(68) & 7.86(50) & 8.42(78) & 7.84(62) \\   
      M\&P \cite{Morningstar:1999rf}  & 6.33(13)   & 6.50(51) & 7.55(11)   & 7.18(11)   \\
      \hline
    FP action                         &[6.74(42)]& & [8.00(35)]&[7.93(78)] \\
    AICP                              & 6.46(18) & 7.66(71) & 8.75(26) & 9.45(71) \\
    \hline
    \end{tabular*}
    \caption{{}Comparison of glueball masses in units of $r_0$. 
      Values in brackets denote masses obtained at a lattice spacing
      $a=0.10$ fm and are not extrapolated to the continuum.}
    \label{tab:cmpgbcont2}
  \end{center}
\end{table}

Table \ref{tab:cmpgbcont} and  \ref{tab:cmpgbcont2} compares our continuum glueball masses obtained with
the anisotropic classically perfect action (AICP) to the results of the
isotropic FP action as well as to results obtained by other collaborations.
There is reasonable agreement between the different determinations of the $0^{++}$,
$2^{++}$, ${0^{++}}^*$, $0^{-+}$, $2^{-+}$ glueball masses. The situation for
the $1^{+-}$ state is different, however. 
This is the heaviest state that occurs in
our analysis and correspondingly difficult to determine on the lattices
used, having a rather small anisotropy $\xi=2$.
Thinking of all the possible sources of errors stated above, which
are (partly) also present in the analyses of the other groups, we tend to
explain this discrepancy with underestimated (or disregarded) systematic
errors as discussed previously. In particular, by looking at the
finite lattice spacing results for $T_1^{+-}$ in Table
(\ref{tab:gb_r0_results}), we observe that the masses determined on the coarser
(and larger) lattices at $\beta=3.15, 3.30$ coincide with the (continuum)
results obtained by other groups and do not even show a significant
discrepancy, despite the considerable difference between the lattice spacings.
The mass determined on the fine (and small) lattice at $\beta=3.50$ however,
is much higher.  This tendency is even amplified by the continuum
extrapolation.  We suspect that this state exhibits strong finite-size effects
pushing up its mass in small volumes.

\begin{figure}[htbp]
\begin{center}
\includegraphics[width=12cm]{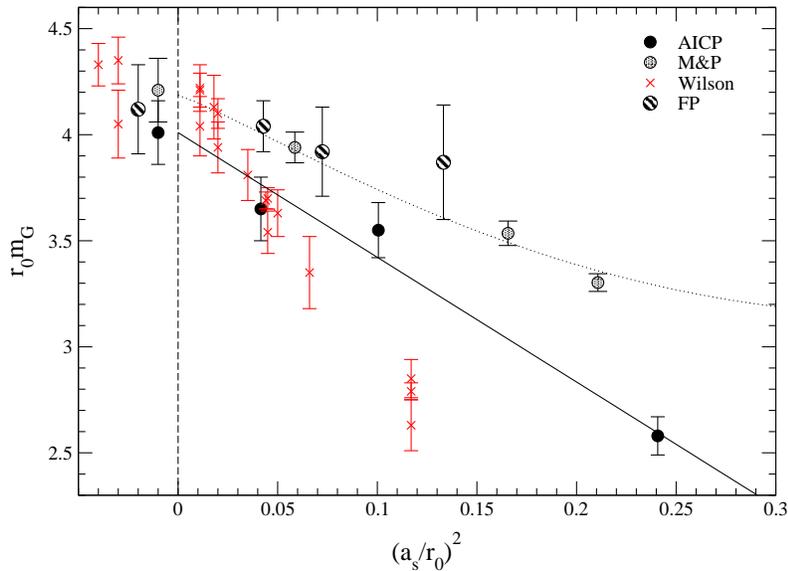}
\end{center}
\caption{{} Lattice results and continuum extrapolation of the
scalar ($0^{++}$) glueball mass for the anisotropic classically
perfect action,
together with results obtained with
different other actions. The solid and dotted lines are the continuum
extrapolations of the AICP and M\&P data, respectively. The different continuum
values for the Wilson action stem from different groups.}
\label{fig:scal_0pp}
\end{figure}

\begin{figure}[htbp]
\begin{center}
\includegraphics[width=12cm]{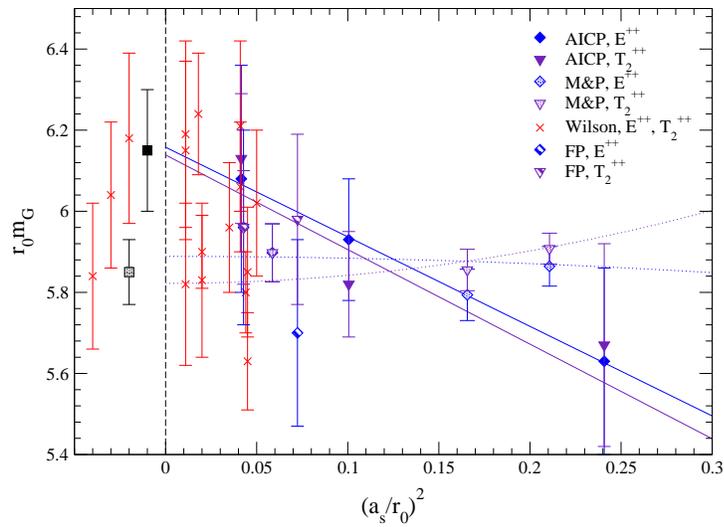}
\end{center}
\caption{{} Lattice results and continuum extrapolation of the
tensor ($2^{++}$) glueball mass for the anisotropic classically
perfect action,
together with results obtained from
simulations employing different other actions. The continuum extrapolations
of the $E^{++}$ and $T_2^{++}$ representations are averaged in order
to get the mass of the single continuum glueball. The solid and dotted lines are the continuum
extrapolations of the AICP and M\&P data, respectively. The different
continuum values for the Wilson action stem from different groups.}
\label{fig:scal_2pp}
\end{figure}

Figures \ref{fig:scal_0pp} and \ref{fig:scal_2pp} compare our
measurements of the $A_1^{++}$, $E^{++}$ and $T_2^{++}$ states to
measurements obtained by other groups, using different actions, as
well as to the isotropic FP action results. Concerning the scalar
glueball, the artifacts of the isotropic FP action at moderate lattice
spacings corresponding to $(a_s/r_0)^2\lesssim 0.15$ ($a_s \lesssim
0.19$~fm) seem to be rather smaller than the ones of the tadpole and
tree-level improved action (M\&P) and certainly much smaller than the
artifacts of the Wilson action. For the AICP action, however, the
situation is less clear. The mass obtained on the coarsest lattice at
$\beta=3.15$ (corresponding to $a_s\approx 0.24$~fm) exhibits large
cut-off effects of about 35\% (compared to about 20\% for the M\&P
action), probably due to the scalar dip.  Concerning the tensor
glueball, the picture is not clear, mainly due to the considerable
statistical errors of the measurements, except for the ones obtained
using the tadpole and tree-level improved anisotropic action.

\section{Repeating the Spatial Blocking Step} \label{ch:xi4act}

As noted in section \ref{ch:construction}, the spatial blocking step
used to obtain a classically perfect $\xi=2$ action may be repeated
straightforwardly in order to generate  actions with higher anisotropies.  Using
another spatial scale 2 blocking step we create a classically perfect
$\xi=4$ gauge action and parametrise it in the same way as the $\xi=2$
action. Furthermore, we measure the renormalised anisotropy and show
that it is indeed feasible to construct actions suitable for MC
simulations of heavy states such as glueballs. The classically
perfect $\xi=4$ action is not yet examined thoroughly, however the
parametrisation is ready (see Appendix \ref{app:xi4act}) and the
analyses of the $\xi=2$ action (see section \ref{ch:xi2act}) may be
repeated for $\xi=4$.

\subsection{Construction}

Each application of the spatial blocking step (slightly) renormalises
the ani\-so\-tro\-py of the action; the renormalised anisotropies of
the final actions thus have to be measured in the end.  However, this
is necessary anyhow if one is interested in comparing most results (other
than e.g.~mass ratios, where the renormalisation of the anisotropy
cancels out) to other collaborations or to the experiment.

In order to be able to repeat the spatial blocking step getting from
\mbox{$\xi=2$} to \mbox{$\xi=4$} we construct a $\xi=2$ action which
is valid on minimised \mbox{$\xi=4$} configurations including into the
fit configurations at different values of $\beta$ over a large
range. The $\xi=2$ action presented in section \ref{ch:xi2act} is not
suitable for this task as it is dedicated to be used solely on largely
fluctuating configurations around $\beta=3.0$. The construction and
the parameters of a suitable intermediate $\xi=2$ action are described in
Appendix \ref{app:xi2intact}.

Once the fine action to be used on the r.h.s. of the renormalisation
group equation, eq.~(\ref{eq:FP_equation}), is ready, we may proceed
along the same lines as for the $\xi=2$ action. We perform the non-linear fit
on 2 $\xi_{\text{ad-hoc}}=6$ configurations each at
$\beta_{\text{ad-hoc}}=$ 4.0, 3.5, 3.0, 2.5, 2.0. It turns out that in
order to obtain parametrisations that are free of dangerous ``traps''
in the $u$-$w$ plane (see Section \ref{sec:xi2constr}) we have to
include the condition $\mathcal{A}(u,w)>0$ at ten points $(u,w)$ into
the non-linear fit. After studying the values of $\chi^2$ as well as the
linear behaviour of the parametrisations we decide to use a set with
$\max(k+l)_{\text{sp}}=3$ and $\max(k+l)_{\text{tm}}=2$. The parameters
are given in Appendix \ref{app:xi4act}.

\subsection{The Renormalised Anisotropy} \label{sec:xi4r}

To check whether the construction of the classically perfect $\xi=4$
action works well and whether the parametrisation reproduces the
input anisotropy, we measure the renormalised anisotropy at one value
of $\beta=3.0$, using the torelon dispersion relation, following the
method described in Section \ref{sec:xir}.  The simulation parameters
are collected in Table \ref{tab:xi4tordet}.

\begin{table}[htbp]
\renewcommand{\arraystretch}{1.3}
  \begin{center}
    \begin{tabular*}{\textwidth}[c]{c@{\extracolsep{\fill}}cc}
      \hline\vspace{-0.05cm} 
      $\beta$ & $S^2\times L\times T$ & \# sweeps / measurements\\
      \hline
      3.0 & $8^2\times 4\times 32$ & 54000 / 10800\\
      \hline
    \end{tabular*}
    \caption{Run parameters for the torelon measurements using the
      $\xi=4$ perfect action. The lattice extension in torelon
      direction $L$, the extension in the two transversal spatial
      directions $S$ as well as the temporal extension $L$ are given.}
    \label{tab:xi4tordet}
  \end{center}
\end{table}

Due to the (expected) coarser spatial lattice spacing we do not need
as many smearing steps as for the $\xi=2$ action and thus perform our
measurements on smearing levels $\mathcal{S}_n, n =
2, 4, 6, 8, 10$ keeping $\lambda_s=0.1$ fixed.

\begin{figure}[htbp]
\begin{center}
\includegraphics[width=12cm]{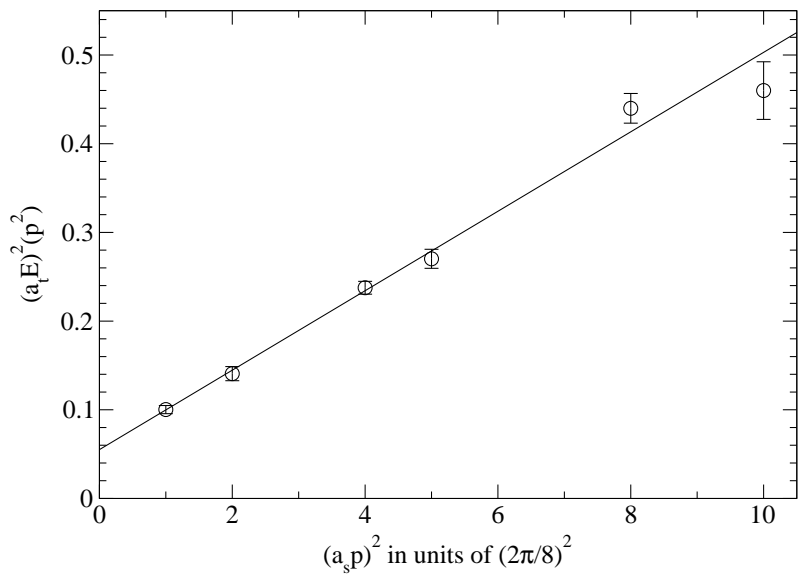}
\end{center}
\caption{Torelon dispersion relation for $\beta=3.0$. The straight line is the correlated fit to $E^2(p)=m_T^2+p^2$ in the range $p^2=1..5$.}
\label{fig:tordisp_xi4_b30}
\end{figure}

Figure \ref{fig:tordisp_xi4_b30} displays the dispersion relation so obtained,
while Table \ref{tab:xi4tor} collects the measured energies $E(p^2)$
determined using variational methods (see Appendix B of
\cite{Niedermayer:2000yx}), together with the number of operators used in the
variational method, the fit ranges and the values of $\chi^2$ per degree of
freedom, $\chi^2/N_{\text{DF}}$.  The renormalised anisotropy as well as the
torelon mass are determined in exactly the same way as for the $\xi=2$ action.
The results are listed in Table \ref{tab:xi4torres}. Furthermore, we may again
evaluate an estimate of the lattice scale using the string picture relation,
eq.~(\ref{eq:finite_size_torelon}), yielding $\sqrt{\sigma}a_s=0.532(16)$, and
thus $r_0/a_s=2.241(86)$ which corresponds to $a_s=0.22$~fm, $a_t=0.060$~fm.
Having a torelon of length $La_s\approx 0.89$~fm the error of the scale
estimate is expected to be about 5-10\% (see Table \ref{tab:xi2scalcomp}).

\begin{table}[htbp]
\renewcommand{\arraystretch}{1.3}
  \begin{center}
    \begin{tabular*}{\textwidth}[c]{c@{\extracolsep{\fill}}cccccccc}
      \hline\vspace{-0.05cm}
      $\beta$ & fit range & $\xi_R$ & $m_{\text{T}} a_t$ & $\chi^2/N_{\text{DF}}$\\
      \hline
      3.0 & 1..5 & 3.71(8) & 0.235(13) & 0.82\\
      \hline
    \end{tabular*}
    \caption{Results of the torelon simulations using the $\xi=4$ perfect action. The fit
      range in $p^2$ is given in units of $(2\pi/S)^2$.}
    \label{tab:xi4torres}
  \end{center}
\end{table}

Concerning the repeated application of the spatial blocking method presented in
this work to yield classically perfect gauge actions for higher
anisotropies, we may conclude that the construction including the
blocking as well as the parametrisation works in exactly the same way
as the construction of the $\xi=2$ action. The resulting action shows a renormalisation
of the anisotropy of about 7\% at the value of $\beta=3.00$
(corresponding roughly to $a_s \simeq 0.2$ fm). Calculations of physical observables such as the
static quark-antiquark potential or glueball masses using the $\xi=4$
classically perfect action are presently absent.

\section{Conclusions} \label{ch:Conclusions}

In this work, we have presented the construction and parametrisation of a
classically perfect anisotropic SU(3) gauge action based on the Fixed Point
Action technique. The recently parametrised isotropic FP action and its
parametrisation ansatz using mixed polynomials of plaquettes built from single
gauge links as well as from smeared links have been the starting point.
Performing one and two purely spatial blocking (RG) steps, respectively,
starting from the isotropic action, and adapting the parametrisation
appropriately, we have obtained parametrised classically perfect actions with
anisotropies $\xi=2$ and 4. 
The $\xi=2$ action has been tested extensively in
measurements of the torelon dispersion relation, of the static quark-antiquark
potential, of the deconfining phase transition as well as of the low-lying
glueball spectrum of pure gauge theory.


The results of the torelon measurements show that the renormalisation
of the anisotropy (due to quantum corrections and parametrisation
artifacts) is small (below 10\% for $\xi=2$ and 4). The rotational
invariance of the action has been examined by measuring the static
quark-antiquark potential, including separations corresponding to a
large set of lattice vectors, again with reassuring results. The glueball
measurements, including some rather heavy states, confirm that the use
of anisotropic lattices facilitates spectroscopy when heavy states are
present. Compared to the isotropic simulations, with the same amount
of computational work it is possible to resolve states with
considerably larger energy, allowing continuum extrapolations
from larger ranges of the lattice spacing. The mass of the scalar
glueball with quantum numbers $J^{PC}=0^{++}$ is measured to be
comparable (1580(60)~MeV) to masses obtained by
other groups (around 1670~MeV); however it shows large cut-off
effects when measured on a lattice with spatial lattice spacing
$a_s\approx 0.25$~fm. This could be caused by a large sensitivity of
the action to a critical point in the fundamental-adjoint coupling
plane, due to adjoint terms in the action. Additionally, one has to be aware
of the manifold sources of
possible errors in measurements of the glueball spectrum. Having at hand the
statistics reached so far, no ultimate statement about the goodness of the
parametrised classically perfect actions can be made yet.

During our studies of both isotropic and anisotropic classically perfect
actions, we have noticed that the examination of scaling properties of
lattice gauge actions is a very delicate problem: Quantities which can
be reliably measured and which are not very sensitive to systematic
factors (like the volume, other states
mixing with the observed ones or the method used for extracting the mass
etc.), e.g.~the critical temperature or the hadronic scale $r_0$
(for moderate lattice spacings) generally exhibit rather small cut-off
effects and thus demand very large statistics, if the differences
between the actions should be explicitly identified.  On the other
hand, some glueball states show large cut-off effects which makes them
interesting objects to study scaling
violations. As pointed out in this work, however, there are a lot of
systematic factors making the extraction of lattice artifacts
difficult.  The task of comparing different actions is even more
difficult if the measurements and analyses are performed independently
by different groups as this introduces additional systematic
discrepancies and ambiguities.

Let us therefore conclude that despite the rich parametrisation which
allows for all the beautiful properties of classically perfect actions
like scale invariance of instanton solutions or rotational invariance,
there is no conclusive evidence of the parametrised classically
perfect actions (isotropic as well as anisotropic) behaving
significantly better than other improved actions (such as the tadpole
and tree-level improved anisotropic gauge action) --- however, there
is also no evidence for the converse. Certainly, one has to consider
the large overhead of the parametrisation as compared to other actions
if one is about to plan pure gauge simulations.

For the future, very accurate scaling tests comparing the classically perfect
actions to other improved gauge actions are desirable. On one hand, these can
include large statistics measurements of the critical temperature and the
hadronic scale, on the other hand, one might perform extensive simulations of
the glueball spectrum, including all the limits to be taken and systematically
excluding all known sources of errors.

With respect to the glueball measurements another promising plan would
be to gain control about the adjoint operators present in the
parametrisation of the classically perfect actions in order to
circumvent the critical point in the adjoint coupling plane such that
the influence of the scalar dip is minimised, similarly to what has
been done for the tree-level and tadpole improved actions by
Morningstar and Peardon \cite{Morningstar:1999dh}.

{\bf Acknowledgements}: We would like to thank Ferenc Niedermayer and
Peter Hasenfratz for valuable discussions. Further thanks go to Simon
Hauswirth, Thomas J\"org and Michael Marti for computer
support. U.W.~acknowledges support by a PPARC special grant.

\begin{appendix}
\section{Action Parameters} \label{app:actions}

In this Appendix we collect the parameters of the FP and the classically
perfect actions that have been used or constructed throughout this
work. The isotropic FP action for coarse configurations has been
presented already in \cite{Niedermayer:2000yx}.  The isotropic action
in Section \ref{app:isointact} is an
intermediate parametrisation valid on configurations typical for MC
simulations that are minimised once, parametrised during the cascade
process for the isotropic action. It is used in the spatial blocking
(see Section \ref{sec:xi2constr}) to minimise the coarse $\xi=2$
configurations, constructed using the ad-hoc anisotropic action
presented in Section \ref{app:adhoc}. The resulting $\xi=2$ perfect
action for coarse configurations is presented in Section
\ref{app:xi2act}.

To repeat the blocking step, we need a $\xi=2$ action which is valid
for $\xi=4$ configurations minimised once in a purely spatial blocking
step. This action is described in Section
\ref{app:xi2intact}. The resulting $\xi=4$ action, finally, is presented in
Section \ref{app:xi4act}.

\subsection{The Intermediate Isotropic Action}\label{app:isointact}
The intermediate isotropic action has been parametrised during the
construction of the isotropic FP action. It is supposed to be valid on
configurations that are obtained by minimising configurations typical
for MC simulations.  It fulfills the $O(a^2)$ Symanzik conditions
(see \cite{Wenger:2000aa,Niedermayer:2000yx}) and uses 
polynomials in the fluctuation parameter $x_\mu(n)$, thence it smoothly approaches the continuum
limit and is expected to interpolate between the rather
coarse configurations mentioned above and the smooth limit. It is not
intended to be used in MC simulations, since its linear behaviour (see
Section \ref{sec:xi2constr}) is not checked.

This action has been used for the (spatial) minimisation of coarse
$\xi=2$ configurations, i.e.~to describe the isotropic configurations on the
r.h.s. of eq.~(\ref{eq:FP_equation}).

The non-linear parameters describing polynomials of order 3 are
(cf.~formula (\ref{eq:smlconstr}), (\ref{eq:asym1}) and
(\ref{eq:eta_x}), (\ref{eq:c_x})):

\begin{small}
\begin{tabular}{llll}
$\eta^{(0)}=0.082$,& $\eta^{(1)}=0.292353$,& $\eta^{(2)}=0.115237$,& $\eta^{(3)}=0.011456$,\\
$c_1^{(0)}=0.282$, & $c_1^{(1)}=-0.302295$, & $c_1^{(2)}=-0.302079$, & $c_1^{(3)}=-0.052309$,\\
$c_2^{(0)}=0.054$, & $c_2^{(1)}=0.298882$, & $c_2^{(2)}=-0.081365$, & $c_2^{(3)}=-0.023762$,\\
$c_3^{(0)}=-0.201671$, & $c_3^{(1)}=0.022406$, & $c_3^{(2)}=0.004090$, & $c_3^{(3)}=0.014886$,\\
$c_4^{(0)}=-0.008977$, & $c_4^{(1)}=0.245363$, & $c_4^{(2)}=0.140016$, & $c_4^{(3)}=0.028783$.\\
\end{tabular}
\end{small}

The linear parameters are collected in Table \ref{tab:isointact}.

\begin{table}[htbp]
\renewcommand{\arraystretch}{1.3}
  \begin{center}
    \begin{tabular}{p{1.2cm}|rrrrr}
\hline
$p_{kl}$     & $l=0$ & $l=1$  & $l=2$ & $l=3$ & $l=4$ \\
\hline
 $k=0$  &            &  0.629227 & -0.556304 &  0.186662 & -0.010110 \\
 $k=1$  & -0.368095  &  0.852428 & -0.199034 &  0.031614 &           \\
 $k=2$  &  0.389292  & -0.207378 & -0.010898 &           &           \\
 $k=3$  & -0.054912  &  0.039059 &           &           &           \\
 $k=4$  & -0.000424  &           &           &           &           \\
\hline
  \end{tabular}
    \caption{{}The linear parameters $p_{kl}$ of the parametrised intermediate isotropic FP action.}
    \label{tab:isointact}
  \end{center}
\end{table}

\subsection{``Ad-hoc'' Anisotropic Actions}\label{app:adhoc}

In the spatial blocking
procedure described in section \ref{ch:construction} one needs coarse
anisotropic ($\xi=$~2, 4, 6, $\ldots$) gauge configurations which are
spatially minimised leading to $\xi'=\xi/2$ configurations. As the
goal of this step is to obtain a perfect anisotropic action with
anisotropy $\xi$ these coarse anisotropic configurations have to be
produced using some other action that is already present. This
requirement might seem to endanger the whole ansatz, however it is not
crucial how the coarse configurations exactly look like as the
perfectness of the resulting coarse action comes from the perfect
action on the fine configuration as well as from the exactness of the
RG transformation. Still, we try to create coarse configurations that
might look similar to future ensembles produced using the perfect
anisotropic action on the coarse level and whose minimised
configurations appear to have an anisotropy approximately $\xi'=\xi/2$.

In order to achieve this, we modify the isotropic FP action by adding a term
$(\xi^2-1)p_{10}^{st}$ (where $p_{10}^{st}$ denotes the simple temporal
plaquette). This modification turns the isotropic Wilson action into the
Wilson action with bare anisotropy $\xi$ and is expected to work approximately
also for our FP action.  The main argument of using this coarse action and not
e.g.~the anisotropic Wilson action is that due to the spatial lattice spacing
$a_s$ being larger than the temporal spacing $a_t$, the $O(a_s^2)$ artifacts
are also expected to be larger than the $O(a_t^2)$ effects. The modification
described above should (approximately) preserve the FP properties
corresponding to $a_s$ and is thus expected to be considerably better than the
naive anisotropic Wilson action.

Using this ad-hoc modification of the isotropic FP action (``ad-hoc''
action) in MC simulations shows (as expected) that (at least for small
anisotropies $\xi\lesssim 5$) its properties on the glueball spectrum
are not comparable with the ones of the isotropic FP action, however
the generated ensembles resemble to the ones generated with true
perfect anisotropic actions much more than ensembles generated with the
Wilson action.

The anisotropies $\xi_{\text{ad-hoc}}$ that have to be used to
generate coarse anisotropic configurations turning into minimised
configurations with anisotropy $\xi'$ are $\xi_{\text{ad-hoc}}\approx
3.2$ for $\xi'=1$ and $\xi_{\text{ad-hoc}}\approx 6$ for
$\xi'=2$. However, this value of $\xi_{\text{ad-hoc}}$ varies
considerably with $\beta$ --- but as stated at the beginning
of this section, the exact form of the coarse configurations is not
essential.

\subsection{The $\mathbf{\xi=2}$ Perfect Action} \label{app:xi2act}

The $\xi=2$ perfect action uses the parametrisation described in
Section \ref{sec:par}.  The number of non-zero asymmetry values
$\eta_i^{(0)}$ is 4, the parameters $c_i^{(0)}$ ($i=1,\ldots,3$) are
splitted into 3 parameters depending on the contribution to the
smeared plaquette. The linear parameters $p_{kl}$ are non-zero for
$0<k+l\leq 4$ for spatial plaquettes and $0<k+l\leq 3$ for temporal
plaquettes.

The non-linear parameters (constants in $x_\mu$) have the values

\begin{small}
\begin{tabular}{llll}
$\eta_1=-0.866007$,& $\eta_2=-0.884110$,& $\eta_3=2.212499$,& $\eta_4=1.141177$;\\
$c_{11}=0.399669$, & $c_{12}=0.519037$, & $c_{13}=-0.071334$, &\\
$c_{21}=-0.076357$,& $c_{22}=-0.031051$,& $c_{23}=-0.282800$, &\\
$c_{31}= 0.032396$,& $c_{32}=-0.015844$,& $c_{33}= -0.046302$.&\\
\end{tabular}
\end{small}

The linear parameters are collected in Table \ref{tab:xi2act_linpar}

\begin{table}[htbp]
\renewcommand{\arraystretch}{1.3}
    \begin{tabular}{p{1.2cm}|rrrrr}
\hline\vspace{-0.05cm}
$p_{kl}^{ss}$     & $l=0$ & $l=1$  & $l=2$ & $l=3$ & $l=4$\\
\hline
 $k=0$  &            &  0.433417 & 0.098921 & -0.116251 & 0.023295\\
 $k=1$  &  0.217599  & -0.272668 & 0.248188 & -0.045278 &         \\
 $k=2$  &  0.316145  & -0.180982 & 0.028817          &  &         \\
 $k=3$  & -0.039521  & 0.003858  &        &          &          \\
 $k=4$  & 0.005443 & & & &\\
\hline
  \end{tabular}

    \begin{tabular}{p{1.2cm}|rrrr}
$p_{kl}^{st}$     & $l=0$ & $l=1$  & $l=2$ & $l=3$\\
\hline
 $k=0$  &            & -0.190195 & 0.554426 & -0.121766\\
 $k=1$  &  1.521212  & -0.328305 & 0.086655 &          \\
 $k=2$  &  0.011178  &  0.020932 &          &          \\
 $k=3$  &  0.022856  &           &          &          \\
\hline
  \end{tabular}

    \caption{The linear parameters of the $\xi=2$ parametrised classically perfect action.}
    \label{tab:xi2act_linpar}
\end{table}

\subsection{The $\mathbf{\xi=2}$ Intermediate Action} \label{app:xi2intact}

In order to be able to repeat the spatial blocking step constructing a
$\xi\approx 4$ action based on the $\xi=2$ perfect action we need a
parametrisation of the $\xi=2$ action which is valid on ($\xi\approx
2$) configurations that are obtained by spatially minimising coarse
$\xi=4$ configurations once. To construct such an action, we perform a
non-linear fit to the derivatives of 5 sets of two configurations each
at $\beta_{\text{ad-hoc}}=$ 6, 10, 20, 50, 100. The non-linear parameters are chosen
to be linear in the fluctuation parameter $x_\mu(n)$. Having four
different parameters $\eta$ and splitting up $c_i$ into three
parameters (as it is done for all anisotropic parametrisations), this
makes 20 non-linear parameters to be fitted which is quite at the edge
of what is still possible on our computers and that is why we restrict
the total number of configurations to 10. Rough checks performed on a
larger number of configurations, with an even larger number of
parameters show however, that the resulting non-linear parameters are
stable and describe the data accurately.

The derivatives and action values of 5 sets of 10 configurations each
are included in the linear fit (where the relative weight of the
action values is chosen to be $1.9\cdot 10^{-2}$ for the
configurations at $\beta_{\text{ad-hoc}}=$ 50, 100 and $7.6\cdot
10^{-4}$ at $\beta=$ 6, 10, 20). A linear set where the parameters
$p_{kl}$ are non-zero for $0<k+l\leq 3$ for spatial plaquettes and
$0<k+l\leq 4$ for temporal plaquettes describes the full action very
well concerning this data. Again, this parametrisation is not intended
to be used in MC simulations, thus the linear behaviour of the action
is not checked.

The non-linear parameters (up to first order in $x_\mu$) have the values 

\begin{small}
\begin{tabular}{llll}
$\eta_1^{(0)}=-1.861267$,& $\eta_1^{(1)}=-0.327466$,& $\eta_2^{(0)}=-1.075610$,& $\eta_2^{(1)}=-0.550398$,\\
$\eta_3^{(0)}= 2.750293$,& $\eta_3^{(1)}= 0.089874$,& $\eta_4^{(0)}= 1.107017$,& $\eta_4^{(1)}= 0.265817$;\\
$c_{11}^{(0)}=0.520960$, & $c_{11}^{(1)}=0.006339$, & $c_{21}^{(0)}=-0.075219$, & $c_{21}^{(1)}=0.059506$\\
$c_{12}^{(0)}=0.266240$, & $c_{12}^{(1)}=0.121035$, & $c_{22}^{(0)}=-0.080771$, & $c_{22}^{(1)}=-0.021515$\\
$c_{13}^{(0)}=0.159372$, & $c_{13}^{(1)}=0.039564$, & $c_{23}^{(0)}=-0.043901$, & $c_{23}^{(1)}=0.009672$\\
\end{tabular}
\end{small}

The linear parameters are collected in Table \ref{tab:xi2intact_linpar}

\begin{table}[htbp]
\renewcommand{\arraystretch}{1.3}
    \begin{tabular}{p{1.2cm}|rrrr}
\hline\vspace{-0.05cm}
$p_{kl}^{ss}$     & $l=0$ & $l=1$  & $l=2$ & $l=3$\\
\hline
 $k=0$  &            &  0.088016 & 0.002225 & -0.000285 \\
 $k=1$  &  0.341850  & -0.015888 &-0.004087 &           \\
 $k=2$  & -0.053007  &  0.010121 &                   &  \\
 $k=3$  &  0.010500  &           &        &             \\
\hline
  \end{tabular}

    \begin{tabular}{p{1.2cm}|rrrrr}
$p_{kl}^{st}$     & $l=0$ & $l=1$  & $l=2$ & $l=3$ & $l=4$\\
\hline
 $k=0$  &            &  0.280043 & 5.077727 &  -13.714872 & 12.739964\\
 $k=1$  &  1.343946  & -6.934825 & 27.673937 & -32.288928 &\\
 $k=2$  &  2.069084  & -17.392027 & 28.248910 &           &\\
 $k=3$  &  3.691733  & -9.584760  &          &           &\\
 $k=4$  &  0.712244  & & & &\\
\hline
  \end{tabular}
    \caption{The linear parameters of the intermediate $\xi=2$ parametrised classically perfect action.}
    \label{tab:xi2intact_linpar}
\end{table}

\subsection{The $\mathbf{\xi=4}$ Perfect Action} \label{app:xi4act}

The $\xi=4$ perfect action uses the parametrisation described in
Section \ref{sec:par}.  The number of non-zero asymmetry values
$\eta_i^{(0)}$ is 4, the parameters $c_i^{(0)}$ ($i=1,\ldots,3$) are
splitted into 3 parameters depending on the contribution to the
smeared plaquette. The linear parameters $p_{kl}$ are non-zero for
$0<k+l\leq 3$ for spatial plaquettes and $0<k+l\leq 2$ for temporal
plaquettes.

The non-linear parameters (constants in $x_\mu$) have the values 

\begin{small}
\begin{tabular}{llll}
$\eta_1=-1.491457$,& $\eta_2=-1.115141$,& $\eta_3=1.510985$,& $\eta_4=7.721347$;\\
$c_{11}=2.014408$, & $c_{12}=0.128768$, & $c_{13}= 0.162296$, &\\
$c_{21}=-0.915620$,& $c_{22}= 0.134445$,& $c_{23}=-0.013383$, &\\
$c_{31}= 1.166289$,& $c_{32}= 0.061278$,& $c_{33}= 0.000759$, &\\
\end{tabular}
\end{small}

The linear parameters are collected in Table \ref{tab:xi4act_linpar}

\begin{table}[htbp]
\renewcommand{\arraystretch}{1.3}
    \begin{tabular}{p{1.2cm}|rrrr}
\hline\vspace{-0.05cm}
$p_{kl}^{ss}$     & $l=0$ & $l=1$  & $l=2$ & $l=3$\\
\hline
 $k=0$  &            &  0.027625 & 0.000052 &  0.000000\\
 $k=1$  &  0.072131  & -0.016852 & -0.000054 &\\
 $k=2$  &  0.036818  &  0.003558 & &\\
 $k=3$  & -0.007413  & & &\\
\hline
  \end{tabular}

    \begin{tabular}{p{1.2cm}|rrr}
$p_{kl}^{st}$     & $l=0$ & $l=1$  & $l=2$\\
\hline
 $k=0$  &            &  0.795779 &  0.621286 \\
 $k=1$  &  2.130563  & -0.286602 & \\
 $k=2$  &  0.076086  & & \\
\hline
  \end{tabular}

    \caption{The linear parameters of the $\xi=4$ parametrised classically perfect action.}
    \label{tab:xi4act_linpar}
\end{table}

\clearpage
\section{Collection of Results} \label{app:rescoll}

\begin{table}[h]
\renewcommand{\arraystretch}{1.3}
  \begin{center}
    \begin{tabular*}{\textwidth}[c]{c@{\extracolsep{\fill}}ccccc}
      \hline\vspace{-0.05cm} 
      $\beta$ & $p^2$ & $M$ & fit range & $a_t E(p^2)$ & $\chi^2/N_{\text{DF}}$\\
      \hline
      3.0  & 0 & 2 & 1--5 & 1.368(30) & 0.19\\
           & 1 & 3 & 1--6 & 1.372(17) & 0.49\\
           & 2 & 3 & 1--6 & 1.451(17) & 0.28\\
           & 4 & 3 & 1--8 & 1.570(21) & 0.46\\
           & 5 & 3 & 1--6 & 1.610(19) & 0.34\\
           & 8 & 2 & 1--3 & 1.707(29) & 0.05\\
      \hline
  3.15$^a$ & 0 & 3 & 2--5 & 0.688(16) & 0.11\\
           & 1 & 4 & 1--5 & 0.806(5)  & 0.06\\
           & 2 & 4 & 1--5 & 0.902(6)  & 0.59\\
           & 4 & 3 & 1--5 & 1.059(8)  & 2.35\\
           & 5 & 4 & 1--5 & 1.151(7)  & 0.16\\
           & 8 & 4 & 1--5 & 1.310(15) & 0.71\\
           & 9 & 4 & 1--5 & 1.356(23) & 1.58\\
           & 10 & 4 & 1--3 & 1.470(17) & 0.42\\
           & 13 & 3 & 1--3 & 1.621(23) & 0.01\\
           & 18 & 3 & 1--3 & 1.762(47) & 0.56\\
      \hline
  3.15$^b$ & 1 & 3 & 1--5 & 1.418(23) & 0.05\\
           & 2 & 3 & 1--5 & 1.541(37) & 0.53\\
           & 4 & 3 & 1--4 & 1.751(67) & 1.20\\
           & 5 & 3 & 1--5 & 1.922(104) & 1.08\\
      \hline
    \end{tabular*}
    \caption{Collection of results of the torelon measurements using the $\xi=2$ perfect action.
For each $\beta$-value and momentum $p^2=p_x^2+p_y^2$ we list the number of operators $M$ kept after
the first truncation in the variational method, the plateau region on which the fit of the correlators
to the form $Z(p^2)\exp(-t E(p^2))$ is performed (fit range), as well as the extracted energy together
with the $\chi^2$ per degree of freedom ($\chi^2/N_{\text{DF}}$). (\emph{Continued in Table \ref{tab:xi2torcoll2}})}
    \label{tab:xi2torcoll}
  \end{center}
\end{table}

\begin{table}[h]
\renewcommand{\arraystretch}{1.4}
  \begin{center}
    \begin{tabular*}{\textwidth}[c]{c@{\extracolsep{\fill}}ccccc}
      \hline\vspace{-0.05cm} 
      $\beta$ & $p^2$ & $M$ & fit range & $a_t E(p^2)$ & $\chi^2/N_{\text{DF}}$\\
      \hline
      3.3  & 0 & 3 & 1--10 & 0.318(6) & 1.39\\
           & 1 & 3 & 1--10 & 0.417(3) & 1.20\\
           & 2 & 3 & 1--5 & 0.493(3) & 0.13\\
           & 4 & 4 & 1--4 & 0.633(4) & 1.26\\
           & 5 & 5 & 1--10 & 0.688(3) & 0.84\\
           & 8 & 5 & 1--8 & 0.830(5) & 1.21\\
           & 9 & 5 & 1--6 & 0.877(6) & 0.42\\
           & 10 & 5 & 1--10 & 0.921(4) & 0.77\\
           & 13 & 5 & 1--10 & 1.036(6) & 0.30\\
           & 18 & 4 & 1--6 & 1.205(12) & 0.15\\
      \hline
   3.5$^a$ & 1 & 5 & 3--8 & 0.305(6) & 0.10\\
           & 2 & 5 & 1--10 & 0.407(4) & 1.65\\
           & 4 & 4 & 1--10 & 0.538(4) & 0.96\\
           & 5 & 5 & 1--10 & 0.586(3) & 1.16\\
           & 8 & 5 & 1--7 & 0.721(4) & 0.09\\
           & 9 & 5 & 2--6 & 0.733(10) & 0.11\\
           & 10 & 4 & 1--8 & 0.795(4) & 0.96\\
           & 13 & 5 & 1--6 & 0.901(4) & 1.26\\
           & 18 & 5 & 1--6 & 1.058(9) & 1.45\\
      \hline
      3.5$^b$ & 1 & 4 & 3--7 & 0.321(5) & 1.00\\
           & 2 & 5 & 1--8 & 0.432(3) & 0.81\\
           & 4 & 5 & 1--8 & 0.586(4) & 0.29\\
           & 5 & 5 & 1--7 & 0.656(3) & 1.24\\
           & 8 & 5 & 2--7 & 0.806(14) & 0.36\\
           & 9 & 5 & 1--7 & 0.859(7) & 0.74\\
           & 10 & 5 & 1--8 & 0.913(4) & 0.68\\
           & 13 & 5 & 2--5 & 1.010(16) & 0.32\\
           & 18 & 3 & 1--5 & 1.215(12) & 0.17\\
      \hline
    \end{tabular*}
    \caption{Collection of results of the torelon measurements using the $\xi=2$ perfect action.
For each $\beta$-value and momentum $p^2=p_x^2+p_y^2$ we list the number of operators $M$ kept after
the first truncation in the variational method, the plateau region on which the fit of the correlators
to the form $Z(p^2)\exp(-t E(p^2))$ is performed (fit range), as well as the extracted energy together
with the $\chi^2$ per degree of freedom ($\chi^2/N_{\text{DF}}$). (\emph{Continuation from Table \ref{tab:xi2torcoll}})}
    \label{tab:xi2torcoll2}
  \end{center}
\end{table}

\begin{table}[h]
\renewcommand{\arraystretch}{1.5}
  \begin{center}
    \begin{tabular*}{\textwidth}[c]{c@{\extracolsep{\fill}}ccccc}
      \hline\vspace{-0.05cm} 
      $\vec{r}$ & $|\vec{r}|$ & $M$ & fit range & $a_t V(\vec{r})$ & $\chi^2/N_{\text{DF}}$\\
      \hline
      (1,0,0) & 1 & 3 & 3--10 & 0.3006(1) & 0.20\\
      (1,1,0) & 1.414 & 3 & 3--10 & 0.3814(2) & 0.21\\
      (1,1,1) & 1.732 & 3 & 3--10 & 0.4250(3) & 0.65\\
      (2,0,0) & 2 & 3 & 6--10 & 0.4469(6) & 0.22\\
      (2,1,0) & 2.236 & 3 & 4--10 & 0.4752(4) & 0.41\\
      (2,1,1) & 2.449 & 3 & 4--10 & 0.4971(5) & 0.83\\
      (2,2,0) & 2.828 & 3 & 4--10 & 0.5309(7) & 0.59\\
      (3,0,0) & 3 & 3 & 4--10 & 0.5445(7) & 1.71\\
      (2,2,1) & 3 & 3 & 5--10 & 0.5456(8) & 0.62\\
      (2,2,2) & 3.464 & 3 & 4--10 & 0.5852(9) & 1.53\\
      (4,0,0) & 4 & 3 & 4--10 & 0.6269(1) & 0.51\\
      (3,3,0) & 4.243 & 3 & 5--10 & 0.6455(16) & 0.94\\
      (4,2,0) & 4.472 & 3 & 5--10 & 0.6630(14) & 0.86\\
      (4,2,2) & 4.899 & 3 & 5--10 & 0.6964(18) & 0.34\\
      (5,0,0) & 5 & 2 & 5--10 & 0.7044(22) & 1.24\\
      (3,3,3) & 5.196 & 3 & 4--10 & 0.7215(17) & 1.31\\
      (4,4,0) & 5.657 & 2 & 6--10 & 0.7509(41) & 1.23\\
      (4,4,2) & 6 & 3 & 5--10 & 0.7783(25) & 0.77\\
      \hline
    \end{tabular*}
    \caption{Collection of results of the off-axis $q\bar{q}$ measurements using the $\xi=2$ perfect
action at $\beta=3.30$. For each separation vector we list the length of the vector in spatial lattice
units, the number of operators $M$ kept after
the first truncation in the variational method, the plateau region on which the fit of the correlators
to the form $Z(\vec{r})\exp(-t E(\vec{r}))$ is performed (fit range), as well as the extracted energy together
with the $\chi^2$ per degree of freedom ($\chi^2/N_{\text{DF}}$).}
    \label{tab:xi2offax}
  \end{center}
\end{table}

\begin{table}[h]
\renewcommand{\arraystretch}{1.5}
  \begin{center}
    \begin{tabular*}{\textwidth}[c]{c@{\extracolsep{\fill}}ccccc}
      \hline\vspace{-0.05cm} 
      $\vec{r}$ & $|\vec{r}|$ & $M$ & fit range & $a_t V(\vec{r})$ & $\chi^2/N_{\text{DF}}$\\
      \hline
      (1,0,0) & 1 & 3 & 2--7 & 0.5053(3) & 0.70\\
      (1,1,0) & 1.414 & 3 & 3--5 & 0.7322(7) & 0.01\\
      (1,1,1) & 1.732 & 3 & 3--8 & 0.8902(14) & 0.60\\
      (2,0,0) & 2 & 3 & 2--8 & 0.9407(10) & 0.89\\
      (2,1,0) & 2.236 & 3 & 2--8 & 1.0642(11) & 0.36\\
      (2,1,1) & 2.449 & 2 & 2--5 & 1.1676(13) & 0.27\\
      (2,2,0) & 2.828 & 3 & 2--6 & 1.3082(18) & 0.17\\
      (3,0,0) & 3 & 3 & 2--6 & 1.3392(25) & 0.16\\
      (2,2,1) & 3 & 3 & 2--5 & 1.3867(20) & 0.26\\
      (2,2,2) & 3.464 & 3 & 2--6 & 1.5675(55) & 0.84\\
      (4,0,0) & 4 & 2 & 1--5 & 1.7429(24) & 0.14\\
      (4,2,0) & 4.472 & 3 & 2--6 & 1.9328(93) & 1.47\\
      (4,2,2) & 4.899 & 3 & 1--4 & 2.1197(34) & 0.15\\
      \hline
    \end{tabular*}
    \caption{Collection of results of the off-axis $q\bar{q}$ measurements using the $\xi=2$ perfect
action at $\beta=3.00$. For each separation vector we list the length of the vector in spatial lattice
units, the number of operators $M$ kept after
the first truncation in the variational method, the plateau region on which the fit of the correlators
to the form $Z(\vec{r})\exp(-t E(\vec{r}))$ is performed (fit range), as well as the extracted energy together
with the $\chi^2$ per degree of freedom ($\chi^2/N_{\text{DF}}$).}
    \label{tab:xi2offax_b300}
  \end{center}
\end{table}

\begin{table}[h]
\renewcommand{\arraystretch}{1.5}
  \begin{center}
    \begin{tabular*}{\textwidth}[c]{c@{\extracolsep{\fill}}ccccc}
      \hline\vspace{-0.05cm} 
      $\beta$ & $r$ & $M$ & fit range & $a_t V(r)$ & $\chi^2/N_{\text{DF}}$\\
      \hline
      3.15 & 1 & 5 & 3--10 & 0.3703(2) & 0.456\\
           & 2 & 5 & 3--7 &  0.6110(4) & 0.735\\
           & 3 & 5 & 3--10 & 0.8019(8) & 1.209\\
           & 4 & 5 & 3--7 & 0.9775(16) & 1.238\\
           & 5 & 5 & 4--9 & 1.129(8)   & 0.519\\
      \hline
      3.50 & 1 & 5 & 3--12 & 0.25910(5) & 1.163\\
           & 2 & 5 & 3--10 & 0.36490(12) & 0.769\\
           & 3 & 5 & 7--12 & 0.42168(38) & 0.664\\
           & 4 & 5 & 5--11 & 0.46587(51) & 0.205\\
           & 5 & 5 & 5--12 & 0.50388(76) & 1.139\\
           & 6 & 5 & 5--11 & 0.54051(96) & 0.853\\
      \hline
    \end{tabular*}
    \caption{Collection of results of the on-axis $q\bar{q}$ measurements using the $\xi=2$ perfect
action at $\beta=$ 3.15, 3.50. For each separation along the axes we list the number of operators $M$ kept after
the first truncation in the variational method, the plateau region on which the fit of the correlators
to the form $Z(r)\exp(-t E(r))$ is performed (fit range), as well as the extracted energy together
with the $\chi^2$ per degree of freedom ($\chi^2/N_{\text{DF}}$).}
    \label{tab:xi2onaxpot}
  \end{center}
\end{table}

\begin{table}[htbp]
  \begin{center}
    \renewcommand{\arraystretch}{1.5}
    \begin{tabular*}{\textwidth}{c@{\extracolsep{\fill}}cccccl}
      \hline
      Channel & $N$ & $t_0/t_1$ & $M$ & fit range &
      $\chi^2/N_{\text{DF}}$ & energies \\
      \hline
      $A_1^{++}$   & 87 & 1/2 & 12 & 1 -- 4 & 0.69 & \bf{0.645(10)} \\
    ${A_1^{++}}^*$ & 59 & 1/2 & 10 & 1 -- 3 & 1.23 & \bf{1.365(104)} \\
      $E^{++}$     & 47 & 1/2 & 7  & 1 -- 3 & 0.10 & \bf{1.405(29)} \\
      ${E^{++}}^*$ & 47 & 1/2 & 5  & 1 -- 3 & 0.17 & \bf{1.985(140)} \\
      $T_2^{++}$   & 47 & 1/2 & 10 & 1 -- 4 & 0.10 & \bf{1.416(32)} \\
    ${T_2^{++}}^*$ & 22 & 1/2 & 6  & 1 -- 3 & 1.20 & \bf{1.990(131)} \\
      $A_2^{++}$   & 21 & 1/2 & 2  & 1 -- 3 & 0.03 & \bf{1.675(147)} \\
      $T_1^{++}$   & 36 & 1/2 & 4  & 1 -- 3 & 0.34 & \bf{1.881(131)} \\
  ${T_1^{++}}^*$   & 36 & 1/2 & 6  & 1 -- 3 & 0.31 & \bf{2.170(182)}\\
      $A_1^{-+}$   & 25 & 1/2 & 3  & 1 -- 3 & 0.001 & \bf{1.588(114)} \\
      $E^{-+}$     & 25 & 1/2 & 2  & 1 -- 3 & 0.01 & \bf{1.849(81)} \\
      ${E^{-+}}^*$ & 25 & 1/2 & 2  & 1 -- 3 & 2.19 & \bf{2.084(268)} \\
      $T_2^{-+}$   & 64 & 1/2 & 4  & 1 -- 3 & 0.02 & \bf{1.850(83)} \\
      $T_1^{+-}$   & 61 & 1/2 & 4  & 1 -- 3 & 0.24 & \bf{1.798(59)} \\
      $T_2^{+-}$   & 18 & 1/2 & 4  & 1 -- 3 & 0.41 & \bf{1.935(177)} \\
    \hline
    \end{tabular*}
    \caption{{}Results from fits to the $\xi=2$, $\beta=3.15$ glueball
      correlators on the $8^3\times 16$ lattice in units of the
      temporal lattice spacing $a_t$: $t_0$/$t_1$ are used in the
      generalised eigenvalue problem, $N$ is the number of initial
      operators measured and $M$ denotes the number of operators kept
      after the truncation in $C(t_0)$.}
    \label{tab:b315_fit_results}
  \end{center}
\end{table}

\begin{table}[htbp]
  \begin{center}
    \renewcommand{\arraystretch}{1.5}
    \begin{tabular*}{\textwidth}{c@{\extracolsep{\fill}}ccccl}
      \hline
      Channel & $t_0/t_1$ & $M$ & fit range &
      $\chi^2/N_{\text{DF}}$ & energies \\
      \hline
      $A_1^{++}$   & 1/2 & 9  & 2 -- 7 & 0.90 & \bf{0.590(17)} \\
    ${A_1^{++}}^*$ & 1/2 & 11 & 1 -- 3 & 1.57 & \bf{1.133(53)} \\
      $E^{++}$     & 1/2 & 11 & 1 -- 5 & 0.97 & \bf{0.983(17)} \\
      ${E^{++}}^*$ & 1/2 & 8  & 1 -- 3 & 0.90 & \bf{1.453(68)} \\
      $T_2^{++}$   & 1/2 & 12 & 1 -- 4 & 1.09 & \bf{0.965(15)} \\
    ${T_2^{++}}^*$ & 1/2 & 10 & 1 -- 4 & 1.68 & \bf{1.386(51)} \\
      $A_2^{++}$   & 1/2 & 4 & 1 -- 3 & 0.01 & \bf{1.511(94)} \\
      $T_1^{++}$   & 1/2 & 8 & 1 -- 5 & 0.84 & \bf{1.492(62)} \\
      $A_1^{-+}$   & 1/2 & 6 & 1 -- 4 & 4.38 & \bf{1.017(39)} \\
      $E^{-+}$     & 1/2 & 3 & 1 -- 4 & 0.74 & \bf{1.365(36)} \\
      $T_2^{-+}$   & 1/2 & 4 & 1 -- 5 & 1.16 & \bf{1.357(25)} \\
      $T_1^{+-}$   & 1/2 & 8 & 1 -- 4 & 0.95 & \bf{1.276(33)} \\
    \hline
    \end{tabular*}
    \caption{{}Results from fits to the $\xi=2$, $\beta=3.3$ glueball correlators on
      the $10^3\times 20$ lattice in units of the temporal lattice spacing $a_t$:
      $t_0$/$t_1$ are used in the generalised eigenvalue
      problem, $M$ denotes the number of operators kept after 
      the truncation in $C(t_0)$.}
    \label{tab:b330_fit_results}
  \end{center}
\end{table}

\begin{table}[htbp]
  \begin{center}
    \renewcommand{\arraystretch}{1.5}
    \begin{tabular*}{\textwidth}{c@{\extracolsep{\fill}}cccccl}
      \hline
      Channel & $N$ & $t_0/t_1$ & $M$ & fit range &
      $\chi^2/N_{\text{DF}}$ & energies \\
      \hline
      $A_1^{++}$   & 91  & 1/2 & 24  & 2 -- 7 & 0.31 & \bf{0.405(13)} \\
    ${A_1^{++}}^*$ & 110 & 0/1 & 110 & 4 -- 6 & 0.30 & \bf{0.720(158)} \\
      $E^{++}$     & 104 & 1/2 & 23  & 2 -- 5 & 0.16 & \bf{0.675(25)} \\
      ${E^{++}}^*$ & 104 & 1/2 & 14  & 1 -- 3 & 0.70 & \bf{1.183(55)} \\
      $T_2^{++}$   & 53  & 1/2 & 19  & 2 -- 5 & 0.31 & \bf{0.681(12)} \\
    ${T_2^{++}}^*$ & 53  & 1/2 & 22  & 1 -- 3 & 0.69 & \bf{1.128(23)} \\
      $A_2^{++}$   & 15  & 1/2 & 6   & 1 -- 3 & 0.05 & \bf{1.212(42)} \\
    ${A_2^{++}}^*$ & 15  & 1/2 & 6   & 1 -- 3 & 1.12 & \bf{1.568(130)} \\
      $T_1^{++}$   & 23  & 1/2 & 11  & 1 -- 4 & 0.05 & \bf{1.245(25)} \\
      $A_1^{-+}$   & 15  & 1/2 & 9   & 1 -- 3 & 0.17 & \bf{0.754(23)} \\
    ${A_1^{-+}}^*$ & 15  & 1/2 & 9   & 1 -- 3 & 0.61 & \bf{1.140(112)} \\
      $E^{-+}$     & 21  & 1/2 & 14  & 1 -- 3 & 0.27 & \bf{0.942(24)} \\
      ${E^{-+}}^*$ & 21  & 1/2 & 11  & 1 -- 3 & 1.20 & \bf{1.406(80)} \\
      $T_2^{-+}$   & 75  & 1/2 & 9   & 2 -- 5 & 0.32 & \bf{0.952(38)} \\
    ${T_2^{-+}}^*$ & 75  & 1/2 & 6   & 1 -- 3 & 1.70 & \bf{1.675(59)} \\
      $T_1^{+-}$   & 53  & 1/2 & 14  & 1 -- 3 & 0.69 & \bf{1.013(17)} \\
      $A_2^{+-}$   & 11  & 1/2 & 5   & 1 -- 3 & 0.07 & \bf{1.224(64)} \\
    ${A_2^{+-}}^*$ & 11  & 1/2 & 5   & 1 -- 3 & 0.79 & \bf{1.726(147)} \\
      $T_2^{+-}$   & 22  & 1/2 & 9   & 2 -- 4 & 0.33 & \bf{1.142(121)} \\
    ${T_2^{+-}}^*$ & 12  & 1/2 & 9   & 1 -- 3 & 1.78 & \bf{1.663(139)} \\
    $E^{+-}$       & 50  & 0/1 & 50  & 1 -- 3 & 1.01 & \bf{1.379(34)} \\
    $T_1^{--}$     & 80  & 0/1 & 80  & 2 -- 4 & 0.57 & \bf{1.200(67)} \\
    $T_2^{--}$     & 80  & 0/1 & 80  & 2 -- 5 & 0.20 & \bf{1.205(81)} \\
    $A_2^{--}$     & 35  & 0/1 & 35  & 1 -- 3 & 0.16 & \bf{1.352(37)} \\
    $A_1^{--}$     & 30  & 0/1 & 30  & 1 -- 4 & 0.14 & \bf{1.703(99)} \\
    \hline
    \end{tabular*}
    \caption{{}Results from fits to the $\xi=2$, $\beta=3.5$ glueball correlators on
      the $12^3\times 24$ lattice in units of the temporal lattice spacing $a_t$:
      $t_0$/$t_1$ are used in the generalised eigenvalue
      problem, $N$ is the number of initial operators measured and $M$ denotes the number of operators kept after 
      the truncation in $C(t_0)$.}
    \label{tab:b350_fit_results}
  \end{center}
\end{table}

\begin{table}[htbp]
  \begin{center}
    \renewcommand{\arraystretch}{1.5}
    \begin{tabular*}{\textwidth}{c@{\extracolsep{\fill}}cccc}
      \hline
      Channel & $J$ & terms in the fit & $\chi^2/N_{\text{DF}}$ & $r_0 m_G$\\
      \hline
      $A_1^{++}$   & 0 & $c$, $c_2$ & 2.72 & \bf{4.01(15)}\\
    ${A_1^{++}}^*$ & 0 & $c$, $c_2$ & 1.43 & \bf{7.66(71)}\\
      $E^{++}$     & 2 & $c$, $c_2$ & 1.06 & \bf{6.16(24)}\\
      ${E^{++}}^*$ & 2 & $c$, $c_2$ & 3.85 & \bf{10.59(63)}\\
      $T_2^{++}$   & 2 & $c$, $c_2$ & 1.98 & \bf{6.14(18)}\\
      $A_2^{++}$   & 3 & $c$, $c_2$ & 1.65 & \bf{11.65(57)}\\
      $A_1^{-+}$   & 0 & $c$        & 2.50 & \bf{6.46(18)}\\
      $E^{-+}$     & 2 & $c$, $c_2$ & 1.06 & \bf{8.73(34)}\\
      $T_2^{-+}$   & 2 & $c$, $c_2$ & 1.07 & \bf{8.77(39)}\\
    \hline
    \end{tabular*}
    \caption{{}Results of the continuum extrapolations of selected glueball representations in terms of the
hadronic scale $r_0$. The continuum spin assignment $J$, the terms in
the fit, constant ($c$) or $(a_s/r_0)^2$ ($c_2$), and
the goodness of the fit, $\chi^2/N_{\text{DF}}$, are also given.}
    \label{tab:cont_ex}
  \end{center}
\end{table}

\begin{table}[htbp]
  \begin{center}
    \renewcommand{\arraystretch}{1.5}
    \begin{tabular*}{\textwidth}{c@{\extracolsep{\fill}}ccccc}
      \hline
      $\Gamma^{PC}$ & $J$ & terms in the fit & $\chi^2/N_{\text{DF}}$ & $m_{\Gamma^{PC}}/m_{T_2^{++}}$ & $r_0 m_G$\\
      \hline
      $E^{++}$     & 2 & $c$ & 1.20 & 1.01(2) & 6.20(30)\\
      ${E^{++}}^*$ & 2 & $c$, $c_2$ & 2.20 & 1.74(12) & 10.68(105)\\
     ${T_2^{++}}^*$ & 2 & $c$, $c_2$ & 3.01 & 1.70(8) & \bf{10.44(80)}\\
      $A_2^{++}$   & 3 & $c$, $c_2$ & 1.11 & 1.90(11) & 11.66(102)\\
      $T_1^{++}$   & 3 & $c$, $c_2$ & 2.60 & 1.90(8) & \bf{11.66(83)}\\
      $A_1^{-+}$   & 0 & $c$        & 1.39 & 1.09(4) & 6.69(44)\\
      $T_2^{-+}$   & 2 & $c$, $c_2$ & 1.23 & 1.45(8) & 8.90(75)\\
      $T_1^{+-}$   & 1 & $c$, $c_2$ & 2.29 & 1.54(7) & \bf{9.45(71)}\\
    \hline
    \end{tabular*}
    \caption{{}Results of the continuum extrapolations of selected glueball ratios, $m_G/m_{T_2^{++}}$. The
 continuum spin assignment $J$, the terms in the fit, constant ($c$)
 or $(a_s/r_0)^2$ ($c_2$), and the goodness of the fit, $\chi^2/N_{\text{DF}}$, are also given. The last
column lists the masses converted to units of $r_0^{-1}$ using the continuum result for the mass of the
tensor glueball $T_2^{++}$, values in bold face will be used further on.}
    \label{tab:cont_ex_ratio}
  \end{center}
\end{table}

\begin{table}[h]
\renewcommand{\arraystretch}{1.5}
  \begin{center}
    \begin{tabular*}{\textwidth}[c]{c@{\extracolsep{\fill}}ccccc}
      \hline\vspace{-0.05cm} 
      $\beta$ & $p^2$ & $M$ & fit range & $a_t E(p^2)$ & $\chi^2/N_{\text{DF}}$\\
      \hline
      3.0  & 1 & 5 & 3--10 & 0.317(5) & 1.34\\
           & 2 & 3 & 4--10 & 0.375(8) & 0.31\\
           & 4 & 3 & 3--11 & 0.487(7) & 0.65\\
           & 5 & 5 & 4--11 & 0.520(11) & 0.35\\
           & 8 & 5 & 3--7 &  0.663(17) & 0.09\\
           & 9 & 3 & 4--11 & 0.754(37) & 0.87\\
           & 10 & 5 & 4--11 & 0.678(32) & 1.56\\
      \hline
    \end{tabular*}
    \caption{Collection of results of the torelon measurement at $\beta=3.0$, using the $\xi=4$ perfect action.
For each momentum $p^2=p_x^2+p_y^2$ we list the number of operators $M$ kept after
the first truncation in the variational method, the plateau region on which the fit of the correlators
to the form $Z(p^2)\exp(-t E(p^2))$ is performed (fit range), as well as the extracted energy together
with the $\chi^2$ per degree of freedom ($\chi^2/N_{\text{DF}}$).}
    \label{tab:xi4tor}
  \end{center}
\end{table}

\end{appendix}

\end{document}